\newcommand{\pa}{\partial}
\newcommand{\be}{\begin{equation}}
\newcommand{\ee}{\end{equation}}
\newcommand{\bea}{\begin{eqnarray}}
\newcommand{\eea}{\end{eqnarray}}
\newcommand{\nn}{\nonumber}

\renewcommand{\a}{\alpha}

\newcommand{\da}{{\dot\alpha}}
\newcommand{\ad}{{\dot a}}
\newcommand{\bd}{{\dot b}}

\renewcommand{\b}{\beta}

\renewcommand{\l}{\lambda}

\newcommand{\q}{\theta}
\newcommand{\bq}{\bar\theta}
\newcommand{\bu}{\bar u}

\newcommand{\ep}{\epsilon}
\newcommand{\m}{\mu}
\newcommand{\n}{\nu}

\newcommand{\cN}{{\cal N}}

\newcommand{\bt}[1]{{\bar t}}
\newcommand{\ts}{\textstyle}

\newcommand{\half}{{\ts \frac{1}{2}}}
\newcommand{\uPu}{P^L_{++}}

\newcommand{\Fg}{\mathcal{F}_g^{(\text{HET})}}
\newcommand{\ua}{\bar{u}_{+1}}
\newcommand{\ub}{\bar{u}_{+2}}
\newcommand{\dc}{\dot{c}}

\newcommand{\uaa}{\bar{u}_{+a}}
\newcommand{\ubb}{\bar{u}_{+b}}

\newcommand{\upp}{\bar{u}_{++}}
\newcommand{\dppsa}{D_{1\dot{1},A}}
\newcommand{\dppsb}{D_{1\dot{1},B}}

\newcommand{\vp}{P^L_{1\dot{1}}}

\newcommand{\vd}{v_{\dot{1}}}

\documentclass[12pt,a4paper]{article}
\input epsf
\usepackage{amsmath,amsfonts,epsfig,color,latexsym}
\usepackage{amssymb}
\usepackage[english, USenglish]{babel}
\usepackage{epsfig}
\usepackage{a4wide}
\usepackage{rotating}
\csname @addtoreset\endcsname{equation}{section}
\author{\\[-0.3cm]\Large I.~Antoniadis\footnote{\tt ignatios.antoniadis@cern.ch}
\footnote{On leave from CPHT (UMR CNRS 7644) Ecole Polytechnique, F-91128 Palaiseau}~, S.~Hohenegger\footnote{{\tt stefan.hohenegger@cern.ch}}~, K.S.~Narain\footnote{{\tt narain@ictp.trieste.it}}~, E.~Sokatchev\footnote{{\tt emeri.sokatchev@cern.ch}}
}
\title{\begin{flushright}{\small CERN-PH-TH/2007-132\\ LAPTH-1201/07}\end{flushright}
\vspace{1cm}\bf{Harmonicity in ${\cal N}=4$ supersymmetry and its quantum anomaly}}
\dateUSenglish
\date{}
\graphicspath{{figures/}}
\begin{document}
\begin{titlepage}
\maketitle
\begin{center}
\renewcommand{\thefootnote}{\fnsymbol{footnote}}\vspace{-0.5cm}
\footnotemark[1]\footnotemark[3]Department of Physics, CERN -- Theory Division\\ CH-1211 Geneva 23, Switzerland\\[0.5cm]
\vspace{-0.3cm}\footnotemark[4]High Energy Section,\\The Abdus Salam International Center for Theoretical Physics,\\Strada Costiera, 11-34014 Trieste, Italy\\[0.5cm]
\vspace{-0.3cm}\footnotemark[5]Laboratoire d'Annecy-le-Vieux de
Physique Th\'{e}orique  LAPTH\footnote[6]{UMR 5108 associ\'{e}e \`{a}
 l'Universit\'{e} de Savoie},\\
B.P. 110,  F-74941 Annecy-le-Vieux, France\\[0.5cm]
\end{center}
\begin{abstract}
The holomorphicity property of ${\cal N}=1$ superpotentials or of ${\cal N}=2$ F-terms involving vector multiplets is generalized to the case of ${\cal N}=4$ 1/2-BPS effective operators defined in harmonic superspace. The resulting {\em harmonicity} equations are shown to control the moduli dependence of the couplings of higher dimensional operators involving powers of the ${\cal N}=4$ Weyl superfield, computed by ${\cal N}=4$ topological amplitudes. These equations can also be derived on the string side, exhibiting an anomaly from world-sheet boundary contributions that leads to recursion relations for the {\em non-analytic} part of the couplings.\\[5pt]
\end{abstract}
\thispagestyle{empty}
\end{titlepage}

\tableofcontents
\pagebreak
\section{Introduction}\label{intro}
An important property of ${\cal N}=2$  F-terms involving vector multiplets is holomorphicity, implying that the corresponding couplings are holomorphic functions of the vector moduli. This applies, for instance, to the couplings $F_g$ of the higher dimensional F-terms $W^{2g}$, where $W$ is the self-dual (chiral) Weyl superfield, appearing in the string effective action~\cite{Antoniadis:1993ze}. On the other hand, the couplings $F_g$ are computed by the topological partition function of the ${\cal N}=2$ twisted Calabi-Yau $\sigma$-model associated to the six-dimensional compactification manifold of type II string theory in four dimensions~\cite{Bershadsky:1993cx, Antoniadis:1993ze}. It turns out, however, that there is a holomorphic anomaly, related to a violation of the conservation of the BRST current in the topological theory, implying that antichiral fields do not decouple at the quantum level~\cite{Cecotti:1992qh, Bershadsky:1993ta, Bershadsky:1993cx}. The anomaly arises from boundary contributions and takes the form of an equation that amounts to a recursion relation for the non-holomorphic part of the couplings $F_g$. From the point of view of the string effective action, it arises from the quantum integration over massless states that is unavoidable when computing on-shell physical amplitudes~\cite{Antoniadis:1993ze}.

These couplings were generalized recently to 1/2-BPS terms of ${\cal N}=4$ compactifications, involving powers of the (superdescendant of the) ${\cal N}=4$ chiral Weyl superfield $K^{++}=D_-D_-W$, where $D_-$ are particular $SU(4)$ projections of the spinor derivatives.  We recall that the ${\cal N}=4$ gravity multiplet contains, besides the graviton and the four gravitini, six graviphoton
s, one complex graviscalar and four spin-1/2 Weyl fermions~\cite{Antoniadis:2006mr}. Moreover, there is an $SU(4)$ R-symmetry, transforming the gravitini in the fundamental and the graviphoton
s in the vector representation. The (linearized on-shell) superfield $K^{++}$ satisfies 1/2-BPS shortening conditions. Its lowest component is the (self-dual) graviphoton field strength and its next bosonic components are the (self-dual) Riemann tensor and the second derivative of the graviscalar. Another basic 1/2-BPS superfield in the ${\cal N}=4$ theory is the (linearized on-shell) vector multiplet $Y^{++}$. Its lowest component are the scalar moduli transforming in the vector representation of $SU(4)$, like the graviphoton field strengths.

In~\cite{Antoniadis:2006mr}  two series of 1/2-BPS couplings were found: ${\cal F}_g^{(1)}{\bar K}^2K^{2g}$ and ${\cal F}_g^{(3)}K^{2(g+1)}$. Here ${\cal F}_g^{(1)}$ and ${\cal F}_g^{(3)}$ are functions of the ${\cal N}=4$ moduli vector multiplets
$Y^{++}$ and of the $SU(4)$ harmonic variables that can again be computed by  topological amplitudes on $K3\times T^2$ of genus $g$ and $g+1$, respectively. Actually, in six dimensions there is also the series $F_g^{(6d)}W_{6d}^{4g}$, where $W_{6d}$ is a similar Weyl superfield of the six-dimensional gravity multiplet and $F_g^{(6d)}$ is given by a topological theory on $K3$~\cite{Berkovits:1994vy}.

In this work, we study the question of what is the analog of ${\cal N}=2$ holomorphicity for such 1/2-BPS  ${\cal N}=4$ couplings. The main novelty of the generalization is that the relevant R-symmetry group becomes non-Abelian, transforming non-trivially the superfields $K^{++}$ and $Y^{++}$. As a consequence, the notion of chirality of the $\cN=2$ theory is replaced by Grassmann analyticity (or 1/2-BPS shortness). The natural framework for studying this problem and covariantizing the expressions is then harmonic superspace~\cite{Galperin:1984av, Galperin:1984bu, Hartwell:1994rp}. By introducing $SU(4)$ harmonic variables one can define $K^{++}$ as a particular harmonic projection of the sixplet of superfields $K_{ij}= -K_{ji} = D_iD_jW$, associated to a corresponding 1/2-BPS subspace of the full ${\cal N}=4$ superspace. Supersymmetry then implies that the coupling coefficients ${\cal F}_g$ are functions of the same harmonic projected vector superfields $Y^{++}$ living in the same 1/2-BPS subspace. Thus,  ${\cal F}_g(Y^{++})$ is independent of the five remaining projections of the sixplet of the scalar moduli. This defines a notion of analyticity that naturally generalizes ${\cal N}=2$ holomorphicity for the chiral $\cN=2$ vector multiplets.

In this work, we show that the above property of analyticity can be formulated in terms of a set of differential constraints on the couplings ${\cal F}_g$ of ${\cal N}=4$ 1/2-BPS effective operators. They express the property of the analytic functions ${\cal F}_g(Y^{++})$ that, when expanded in powers of the harmonic variables and the scalar fields, the coefficients should form symmetric traceless tensors of $SO(6)$. This yields two non-trivial equations. The first requires one scalar and one harmonic derivative to vanish and coincides with the so-called `harmonicity' equation found previously in string computations, up to an anomaly~\cite{Berkovits:1994vy,Ooguri:1995cp,Antoniadis:2006mr}. The second involves two scalar derivatives and gets modified in supergravity by an additive constant term due to the curvature of the scalar kinetic terms. Both equations are checked by an explicit string computation for  ${\cal F}_g^{(3)}$, which receives one-loop corrections on the heterotic side for all $g$, and are found to be corrected by anomaly terms due to world-sheet boundary contributions that spoil the naive expectation of analyticity, in a way similar to the holomorphic anomaly equation of the ${\cal N}=2$ $F_g$'s. The resulting equations are reduced again to recursion relations for the non-analytic part of the moduli-dependent couplings.

We finally extend the above results to six-dimensional $\cN=(1,1)$ supersymmetry, where the R-symmetry group is $SO(4)$. In particular, we consider the decompactification limit of ${\cal F}_g^{(3)}$ that gives rise to a new six-dimensional series of 1/2-BPS couplings of the form ${\cal F}_g^{\rm dec}W_{6d}^{2(g+1)}$. These couplings, although not exactly topological in six dimensions (the space-time part is not decoupled), become topological upon compactification to four dimensions on a two-torus. Despite this fact, ${\cal F}_g^{\rm dec}$ satisfy the same analyticity condition as $F_g^{(6d)}$ of $W_{6d}^{4g}$ since they are both 1/2-BPS. We then derive the corresponding analyticity equations, together with the anomaly terms.

The paper is organized as follows. In Section~\ref{global}, we describe the analyticity conditions of the 1/2-BPS couplings in the case of global ${\cal N}=4$ supersymmetry. We first introduce the $SU(4)$ harmonic variables and the harmonic projected vector and Weyl superfields, $Y^{++}$ and $K^{++}$ respectively. We then derive the differential equations for the couplings ${\cal F}_g$ of the higher-derivative operators involving powers of $(K^{++})^2$, as described above. In Section~\ref{coset}, we study the effects of the curvature of the scalar manifold that parametrizes the coset $SO(6,n)/SO(6)\times SO(n)$, where $n$ is the number of vector multiplets. We show in particular that the second-order derivative equation in the scalar fields gets modified by an additional term proportional to the Weyl weight of the operator. In Section~\ref{css}, we go to curved superspace in the framework of ${\cal N}=4$ conformal supergravity and derive the fully covariantized final expressions of the higher-derivative couplings. In Section~\ref{Sect:Review}, we give a brief review of the ${\cal N}=4$ topological amplitudes in string theory and recall the expression for ${\cal F}_g^{(3)}$ obtained from a one-loop string computation on the heterotic side. In Section~\ref{Sect:FirstOrder}, we derive the harmonicity relation which is first order in the scalar field derivatives, exhibiting a boundary anomaly that invalidates the expected vanishing result. In Section~\ref{Sect:SecondOrderRelation}, we obtain the second-order constraint which is also modified by an anomaly. In Section~\ref{sixdim}, we generalize our analysis to six dimensions. We first introduce the $SO(4)$ harmonic variables and derive the harmonicity equations for the couplings of the 1/2-BPS terms. We then consider the decompactification limit of ${\cal F}_g^{(3)}$ and compute the two analyticity equations modified by the anomalous terms. 
Finally, Section~\ref{Sec:Concl} contains some concluding remarks.

\section{Global $\cN=4$ supersymmetry}\label{global}

\subsection{$SU(4)$ harmonic variables}

We consider $\cN=4$ supersymmetry in four dimensions whose automorphism group is
$SU(4)$. We introduce harmonic variables \cite{Galperin:1984av,Galperin:1984bu,Hartwell:1994rp} on the coset $SU(4)/S(U(2)\times U(2))$ in the
form of matrices $(u^{+a}_i, \, u^{-\ad}_i) \in SU(4)$. They have an index
$i=1\ldots4$ transforming under the fundamental irrep of $SU(4)$ and indices $a, \ad =
1,2$ of $SU(2)\times SU(2)$ as well as $U(1)$ charges $\pm 1$. Together with their
complex conjugates $\bu^i_{+a} = \overline{(u^{+a}_i)}, \, \bu^i_{-\ad} =
\overline{(u^{-\ad}_i)}$  they satisfy the unitarity conditions
\begin{eqnarray}
  &&u^{+a}_i\, \bu^i_{+b} = \delta^a_b\,, \quad u^{-\ad}_i\, \bu^i_{-\bd} = \delta^\ad_\bd \,, \quad u^{+a}_i\, \bu^i_{-\bd} = u^{-\ad}_i\, \bu^i_{+b} = 0 \nn \\
  &&u^{+a}_i\, \bu^j_{+a} + u^{-\ad}_i\, \bu^j_{-\ad} = \delta^j_i\label{12'}
\end{eqnarray}
and the unit determinant condition
\begin{equation}\label{12}
    \ep^{ijkl} u^{+a}_i u^{+b}_j u^{-\ad}_k u^{-\bd}_l = \ep^{ab}\ep^{\ad\bd}
\end{equation}
(with $\ep^{1234} = \ep^{12}= \ep^{\dot 1\dot 2} = -\ep_{12}= -\ep_{\dot 1\dot 2} = 1$).

The harmonic functions have harmonic expansions homogeneous under the action of the
subgroup $S(U(2)\times U(2))$. The harmonic expansions are organized in irreps of
$SU(4)$, keeping the balance of projected indices so that the overall $SU(2)\times
SU(2)$ indices and the $U(1)$ charge are always the same. In what follows we shall often
make use of functions depending on vector-like combinations of $SU(4)$ harmonics (i.e.,
with harmonics on $SO(6)/SO(4)\times SO(2)$) of the type $u^M_{ij} = -u^M_{ji}$, $M= (++,--, a\ad)$ (and their conjugates $\bu_{M}^{ij} = \overline{u^{M}_{ij}}$)
\begin{eqnarray}
  && u^{++}_{ij} = u^{+a}_i \ep_{ab} u^{+b}_j \nn\\
  && u^{--}_{ij} = u^{-\ad}_i \ep_{\ad\bd} u^{-\bd}_j \label{00} \\
  && u^{a\ad}_{ij} = u^{+a}_{[i} u^{-\ad}_{j]}\ ,  \nn
\end{eqnarray}
where $[ij]$ denotes weighted antisymmetrization. They form $SO(6)$ matrices $u^M_N = \Gamma_N^{ij}u^M_{ij}$ where $\Gamma^M$ are the
gamma matrices of $SO(6)$. 
The vector-like harmonics satisfy algebraic conditions expressing the fact that $u^M_N \in SO(6)$ and  following from the conditions on the underlying $SU(4)$ harmonics:
\begin{eqnarray}
  &&u^{++}_{ij} \ep^{ijkl} u^{--}_{kl} = 4 \nn \\
  &&u^{a\ad}_{ij} \ep^{ijkl} u^{b\bd}_{kl} = \ep^{ab}\ep^{\ad\bd}  \label{vchac}\\
   &&u^{++}_{ij} \ep^{ijkl} u^{++}_{kl} = u^{--}_{ij} \ep^{ijkl} u^{--}_{kl} = u^{++}_{ij} \ep^{ijkl} u^{a\ad}_{kl}  = u^{--}_{ij} \ep^{ijkl} u^{a\ad}_{kl}= 0  \nn \\
   &&u^{++}_{ij} u^{--}_{kl} + u^{--}_{ij} u^{++}_{kl} - 2\,  u^{a\ad}_{ij}\ep_{ab} \ep_{\ad\bd} u^{b\bd}_{kl}=   \ep_{ijkl} \nn
\end{eqnarray}
An example of a harmonic function which we shall frequently encounter is $\phi^{++}(u) = \phi^{ij} u^{++}_{ij} + \phi^{ij\, kl}_{mn}
u^{++}_{ij} u^{++}_{kl} u^{mn}_{++} + \cdots$. The first component in this expansion is
a \underline{6} of $SU(4)$ (or a vector of $SO(6)$) $\phi^{ij} = - \phi^{ji}$. The
higher components give rise to higher-dimensional irreps, but we shall not need them
here.

The harmonic derivatives can be viewed as the covariant derivatives on the harmonic coset  $SU(4)/S(U(2)\times U(2))$, or equivalently, as the generators of the $SU(4)$ algebra written down in an $S(U(2)\times U(2))$ basis  (see Section \ref{coset}). This means that they are invariant under the left action of the group $SU(4)$, but covariant under the right action of the subgroup $S(U(2)\times U(2))$. They can be split into generators of the subalgebra $S(U(2)\times U(2))$:
\begin{eqnarray}
  D_{+a}{}^{+b} &=& \bigl{(} u^{+b}_i \frac{\pa}{\pa u^{+a}_i } - \bu_{+a}^i \frac{\pa}{\pa \bu_{+b}^i}\bigr{)} - \mbox{trace} \nn\\
  D_{-\ad}{}^{-\bd} &=& \bigl{(}u^{-\bd}_i \frac{\pa}{\pa u^{-\ad}_i } - \bu_{-\ad}^i \frac{\pa}{\pa \bu_{-\bd}^i}\bigr{)} - \mbox{trace} \label{subhd}\\
  D_0 &=& \bigl{(}u^{+a}_i \frac{\pa}{\pa u^{+a}_i } - \bu_{+a}^i \frac{\pa}{\pa \bu_{+a}^i} \bigr{)}- \bigl{(}u^{-\ad}_i \frac{\pa}{\pa u^{-\ad}_i } - \bu_{-\ad}^i \frac{\pa}{\pa \bu_{-\ad}^i} \bigr{)}\nn
\end{eqnarray}
and of the coset:
\begin{eqnarray}
  D_{+a}{}^{-\bd} &=& u^{-\bd}_i \frac{\pa}{\pa u^{+a}_i } - \bu_{+a}^i \frac{\pa}{\pa \bu_{-\bd}^i} \nn\\
  D_{-\ad}{}^{+b} &=& u^{+b}_i \frac{\pa}{\pa u^{-\ad}_i } - \bu_{-\ad}^i \frac{\pa}{\pa \bu_{+b}^i} \ . \label{cosethd}
\end{eqnarray}
The harmonic derivatives are differential operators preserving the defining algebraic constraints (\ref{12'}), (\ref{12}).

The derivatives (\ref{subhd}) act homogeneously on the harmonic functions. For instance, the function $\phi^{++}(u)$ above has no $SU(2)\times SU(2)$ indices, but has $U(1)$ charge, hence
\begin{equation}\label{exhf}
    D_{+a}{}^{+b} \phi^{++}(u) = D_{-\ad}{}^{-\bd} \phi^{++}(u) = 0\,, \qquad  D_0  \phi^{++}(u) = 2 \phi^{++}(u)\ .
\end{equation}
The harmonic expansion of this function defines an infinitely reducible representation of $SU(4)$. It can be made irreducible by requiring that the raising operator $D_{-\ad}{}^{+b}$ annihilate the function:
\begin{equation}\label{ropcon}
    D_{-\ad}{}^{+b} \phi^{++}(u) = 0\ \quad \Rightarrow \quad  \phi^{++}(u) = \phi^{ij} u^{++}_{ij} \ .
\end{equation}
In other words, such a function is a highest-weight state of the \underline{6} of $SU(4)$. The irreducibility condition (\ref{ropcon}) is also called a condition for harmonic (H-) analyticity.

\subsection{Grassmann analytic on-shell superfields}

The introduction of harmonic variables allows us to define `1/2 BPS short' or Grassmann
(G-) analytic\footnote{A more systematic derivation of the G-analytic superfields as functions on a coset of the $\cN=4$ superconformal algebra $PSU(2,2/4)$ will be given in Section \ref{css}.} superfields.\footnote{The notion of Grassmann analyticity (with breaking of the R symmetry) was first proposed in \cite{Galperin:1980fg} in the context of the $\cN=2$ hypermultiplet. Later on it was made R-symmetry covariant in the framework of $\cN=2$ harmonic superspace in \cite{Galperin:1984av}. The harmonic superspace description of the $\cN=3$ {\it off-shell} vector multiplet was given in \cite{Galperin:1984bu}, and that of the $\cN=4$ {\it on-shell} vector multiplet in \cite{Hartwell:1994rp}.}  They depend only on half of the Grassmann variables which can
be chosen to be $\q^{+a}_\a = \q^i_\a\, u_i^{+a}$ and $\bq^\da_{-\ad} =
\bu^i_{-\ad}\,\bq^\da_i $. One such superfield is the {\it linearized  on-shell} vector
multiplet
\begin{eqnarray}
  Y^{++}(x^\mu,\q^+,\bq_-,u) &=& \phi^{ij} u^{++}_{ij} + \q^{+a}_\a\,\ep_{ab}\, \psi^{\a\, i}\,u^{+b}_i +\bu^i_{-\ad}\,\bar\psi_{\da\,i}\, \ep^{\ad\bd}\, \bq^\da_{-\bd} \label{01} \\
  &+& \q^{+a}\sigma^{\m\n}\q^{+b}\ep_{ab}\, F_{(+)\m\n} + \bq_{-\ad}\tilde\sigma^{\m\n}\bq_{-\bd}\ep^{\ad\bd}\, F_{(-)\m\n} + \mbox{derivative terms} \ . \nn
\end{eqnarray}
Here $\phi^{ij}= \frac{1}{2}\ep^{ijkl} \bar\phi_{kl}$ are the six real scalars,
$\psi^\a_i$ are the four Majorana gluinos and $F_{(\pm)\m\n}$ is the (anti)self-dual
part of the gluon field strength. To exhibit manifest G-analyticity, one has to choose the appropriate analytic basis in superspace,
\begin{equation}\label{chbashss}
    x^\mu\ \to \ x^\mu + i \q^{+a}\sigma^\mu\bar\q_{+a} + i \q^{-\ad}\sigma^\mu\bar\q_{-\ad}\ ,
\end{equation}
analogous to the familiar chiral basis. Note that the harmonic dependence here is cut down
to linear in the vector-like and fundamental harmonics. This is typical for on-shell
multiplets which, in addition to the G-analyticity condition, also satisfy the H-analyticity condition
\begin{equation}\label{hansf}
    D_{-\ad}{}^{+b} Y^{++}(\q^+,\bq_-,u) = 0\ .
\end{equation}
Here the harmonic derivative is supersymmetrized by going to the manifestly G-analytic superspace coordinates (\ref{chbashss}). One can show that the `ultrashort' on-shell superfield (\ref{01}) is the solution to the simultaneous conditions for G- and H-analyticity \cite{Hartwell:1994rp,Andrianopoli:1999vr}.

Another example of a G-analytic superfield is the {\it linearized}  on-shell Weyl
multiplet. It is obtained from the off-shell chiral Weyl superfield \cite{Bergshoeff:1980is}
\begin{equation}\label{14}
    W(\q_\a^i) = \Phi + \q_\a^i \q_\b^j (\sigma_{\m\n}^{\a\b} T_{(+)[ij]}^{\m\n} + \ep^{\a\b} S_{(ij)}) + \cdots \ .
\end{equation}
Here $\Phi$ is the physical scalar and $T_{(+)}$ is the self-dual part of the sixplet of
graviphoton field strengths, while $S$ is an auxiliary field. On shell the latter must vanish, hence
the additional constraint
\begin{equation}\label{15}
   \ep_{\a\b} D_i^\a D_j^\b\ W = 0\,.
\end{equation}
Now, define the superfield (a superdescendant of $W$)
\begin{equation}\label{16}
    K^{++}_{\m\n} = (\sigma_{\m\n})_{\a\b}D_{-\ad}^\a D_{-\bd}^\b\ep^{\ad\bd}\ W
\end{equation}
where we have projected the $SU(4)$ indices of $D^\a_i D^\b_j$ with the harmonics
$\bu_{-\ad}^i \bu_{-\bd}^j$ . This superfield is annihilated by half of the spinor
derivatives and hence is 1/2 BPS short. Indeed, this is true for the projections $\bar
D^{+a}_{\da}$ since $\{\bar D^{+a}, D_{-\bd}\}=0$ and $\bar D^i_{\da}W=0$ (chirality).
Further, hitting (\ref{16}) with $D_{-\ad}^\gamma$ we obtain zero as a consequence of
the projection of the on-shell constraint (\ref{15}) with $\bu_{-\ad}^i \bu_{-\bd}^j$.
We conclude that $K^{++}_{\m\n}$ satisfies the G-analyticity constraints
\begin{equation}\label{16'}
    D_{-\ad}^\a K^{++}_{\m\n} = \bar D^{+a}_{\da} K^{++}_{\m\n} = 0
\end{equation}
which imply that it depends on half of the $\q$'s (bosons only):
\begin{equation}\label{17}
    K^{++}_{\m\n}(\q^+,\bq_-,u) = T_{(+)\m\n}^{ij}  u^{++}_{ij}  + \q^{+a}\sigma^{\l\rho}\q^{+b}\ep_{ab}\,  R_{(+)\m\n\l\rho} + \bq_{-\ad}\tilde\sigma^\l \sigma_{\m\n} \sigma^\rho \bq_{-\bd}\ep^{\ad\bd}\,  \pa_\l \pa_\rho \Phi + \cdots\ .
\end{equation}
In addition, the harmonic dependence of $K^{++}_{\m\n}$ is restricted to be linear. As in (\ref{hansf}), this follows from the condition for H-analyticity
\begin{equation}\label{hwconK}
    D_{-\ad}{}^{+b} K^{++}_{\m\n} = 0 \ .
\end{equation}
This is another example of an ultrashort superfield. Note, however, that it is not a primary object but rather a superdescendant of the chiral  on-shell Weyl
multiplet.

Repeating the same steps, but this time starting with the antichiral superfield $\bar
W(\bq)$ we obtain the other half of the on-shell Weyl multiplet. It is again described
by an ultrashort superfield of the same type,
\begin{equation}\label{18}
    \bar K^{++}_{\m\n}(\q^+,\bq_-,u) = T_{(-)\m\n}^{ij}   u^{++}_{ij} + \bq_{-\ad} \tilde\sigma^{\l\rho}\bq_{-\bd}\ep^{\ad\bd}\, R_{(-)\m\n\l\rho} + \q^{+a}\sigma^\l\tilde\sigma_{\m\n} \tilde\sigma^\rho \q^{+b}\ep_{ab} \,   \pa_\l \pa_\rho \bar\Phi + \cdots
\end{equation}
Note that in the $\cN=4$ G-analytic superspace there exists a special conjugation $\
\widetilde{}\ $ combining complex conjugation with a reflection on the harmonic coset,
such that G-analyticity is preserved. In this sense $Y^{++} = \widetilde{Y^{++}}$ and
$\bar K^{++} = \widetilde{K^{++}}$, which implies, in particular, the reality condition
on the six scalars in $Y$.

\subsection{Higher-derivative couplings}

After having defined the G-analytic superfields (\ref{01}) and (\ref{17}), we now want to construct the corresponding effective action couplings. Recently, by studying special higher loop scattering processes in the gravitational sector of type II superstring theory compactified on $K3\times T^2$ (or the corresponding dual formulation of heterotic string on $T^6$, as we will review in Section \ref{Sect:Review}), the following two terms were found in \cite{Antoniadis:2006mr}:
\begin{eqnarray}
  S_1 &=&\int \ d^4x \ du \ d^4\q^+ d^4\bq_{-} \ (\bar K^{++}_{\m\n} \bar K^{++\, \m\n})\, (K^{++}_{\rho\sigma} K^{++\, \rho\sigma})^{g}\, F_1^{-4(g-1)}(Y^{++}_A,u)\ , \label{19}\\
  S_2 &=& \int \ d^4x \ du \ d^4\q^+ d^4\bq_{-} \ (K^{++}_{\m\n} K^{++\, \m\n})^{g+1}\, F_2^{-4(g-1)}(Y^{++}_A,u)\ , \label{19'}
\end{eqnarray}
where  $A=1\cdots n$ is an $SO(n)$ vector index
labeling the coordinates of the coset of physical scalars (see Section \ref{coset}).
In fact, if considered as $g$- and $(g+1)$-loop contributions respectively, both of these terms lead to so-called {\it topological amplitudes}, that is the corresponding physical amplitudes are computed by correlation functions of the $\cN=4$ topological string on $K3\times T^2$. However, unlike the $\cN=2$ case (see \cite{Antoniadis:1993ze}) these correlation functions are not simply the topological partition function, but differ from it by additional operator insertions in the twisted version of the theory.
Actually, for convenience and notational simplicity, we changed the notation of the two couplings from \cite{Antoniadis:2006mr}: $F_1^{-4(g-1)}$ corresponds to ${\cal F}_g^{(1)}$, while $F_2^{-4(g-1)}$ corresponds to ${\cal F}_g^{(3)}$, with the upper index denoting the $U(1)$ charge.

It is important to stress upon two points concerning the effective action terms (\ref{19}) and (\ref{19'}):
\begin{enumerate} 
\item The Grassmann measure is G-analytic, i.e. it involves only half of the projected
$\q$'s, and so must be the integrand, otherwise supersymmetry will be broken. This is
why we have to use the {\it linearized on-shell} superfields $Y^{++}$ and $K^{++}$ which
are G-analytic like the measure.
\item 
The harmonic integral should produce an $SU(4)$ invariant, i.e. it  picks out the
$SU(4)$ singlet part of the integrand. This is only possible if the latter is a {\it
chargeless} harmonic function. For example, $f(u) = f_0 + f^{ij}_{kl}u^{++}_{ij}
u^{kl}_{++} + \cdots $ integrates to $\int du \, f(u) = f_0$, but a charged function
like $f^{++} = f^{ij}u^{++}_{ij}  + \cdots$ will have a vanishing integral.  Notice that
for this reason the harmonic integral should always be done {last}, after the Grassmann
integrals, since the latter are charged.
\end{enumerate}
In our case (\ref{19}), (\ref{19'}) the functions $F_{1,2}$ carry $U(1)$ charge $-4(g-1)$
needed to compensate that of the factor $K$ $(+4(g+1))$ and of the Grassmann measure
$(-8)$. Given the fact that the argument $Y^{++}$ of $F$ has a positive charge, we have
to introduce a set of {\it constant}  $SU(4)$ multispinors
\begin{equation}\label{20'}
     \xi^{-2p}(u) \equiv \xi_{(i_1 \cdots i_p) (j_1 \cdots j_p)}\  \bu_{++}^{i_1 j_1} \cdots \bu_{++}^{i_p j_p} + \ldots
\end{equation}
thus explicitly breaking $SU(4)$.\footnote{The other possibility, which we do not
consider here, would be to use singular functions involving inverse powers of fields.} The dots denote higher-order terms in the harmonic expansion of the coefficients
$\xi(u)$ which will not be of interest for us, see below. Note that the product of vector-like harmonics forms an  irreducible representation of
$SO(6)$, a symmetric traceless tensor of rank $p$ (recall that $(\bu_{++})^2 =0$, see (\ref{vchac})).  In
$SU(4)$ notation this means that the indices $i$ and $j$ of the coefficients $\xi$ are
separately symmetrized, but antisymmetrized between $i$ and $j$, i.e. we are dealing
with the irrep $(0p0)$. In what follows this fact will be of crucial importance. So, we
consider the potential ($m=2(g-1)$; the $SO(n)$ index $A$ and the labels $1,2$ are suppressed)
\begin{equation}\label{20}
    F^{-2m}(Y^{++},u) = \sum_{n=0}^\infty  \ \xi^{{-2(m+n)}}(u) \ (Y^{++})^n\ .
\end{equation}

The factors $K$ in (\ref{19}) and (\ref{19'}) contribute, among others, the terms
\begin{equation}\label{21}
    (\q^+)^4({\bq}_-)^4\ R^2_{(+)} \ R^2_{(-)}\
    (T^{++}_{(+)})^m\,,   \qquad   (\q^+)^4({\bq}_-)^4\ R^2_{(+)}(\pa\pa\Phi)^2 \
    (T^{++}_{(+)})^m\,,
\end{equation}
respectively. The $\q$'s saturate the superspace measure and are integrated out. The
remainder has a harmonic charge,
\begin{equation}\label{22}
        (T^{++})^{m} = T^{(i_1(j_1} \cdots T^{i_{m}) j_{m})}\ u^{++}_{i_1 j_1} \cdots u^{++}_{i_m j_m}
\end{equation}
which is compensated by the factor $F$ in order to have a non-vanishing harmonic
integral (i.e., an $SU(4)$ singlet). Clearly, (\ref{22}) is an irrep of $SO(6)$, a
symmetric traceless tensor of rank $m$. This can be reformulated as the highest-weight condition (cf. (\ref{ropcon}))
\begin{equation}\label{hwcon}
    D_{-\ad}{}^{+b} (T^{++})^{m} = 0 \ .
\end{equation}
A similar condition holds for the entire effective action expressions (\ref{19}) and (\ref{19'}) of the graviphoton field strength superfield.

The singlet needed for the harmonic integral is obtained by combining (\ref{22}) with
the matching irrep in $F$. Consider the harmonic structure of $F$ (all
$\q=0$):
\begin{eqnarray}
  F^{-2m}(\phi^{++},u) &=& \sum_{n=0}^\infty  \ \xi_{(i_1 \cdots i_{m+n}) (j_1 \cdots j_{m+n})}\    \bu_{++}^{i_1 j_1} \cdots \bu_{++}^{i_{m+n} j_{m+n}} \nonumber\\
  & \times&  \phi^{(k_1(l_1} \cdots \phi^{k_{n}) l_{n})}  u^{++}_{k_1 l_1} \cdots u^{++}_{k_n l_n} \ .\label{23}
\end{eqnarray}
Here we have restricted the harmonic expansion (\ref{20'}) of the coefficient function $\xi^{{-2(m+n)}}(u)$ to the lowest-rank $SO(6)$ irrep. The higher-rank terms are irrelevant due to the gauge invariance of
the couplings (\ref{19}), (\ref{19'}). Indeed, consider adding a total supersymmetrized harmonic
derivative $D_{-\ad}{}^{+b} \Lambda^{(-4g+2) }{}_b^{\ad}(\q^+,\bq_-,u)$ to the potential
$F^{-4(g-1)}$. After
integration by parts (the G-analytic measure allows this), $D_{-\ad}{}^{+b}$ annihilates the on-shell superfield $K^{++}$ (recall (\ref{hwconK})), hence the gauge invariance of
(\ref{19}), (\ref{19'}) with the G-analytic parameter $\Lambda$. By examining the harmonic
expansion of $\Lambda(0,0,u)$ one can show that all the omitted terms in (\ref{23}) can be gauged away.

The gauge-fixed function (\ref{23}) satisfies two differential conditions. The first one expresses the fact that it is a function only of the projection $\phi^{++}$ of the $SO(6)$ vector of physical scalars:
\begin{equation}\label{firstcon}
    \frac{\pa}{\pa \phi^{--}} F^{-2m} = \frac{\pa}{\pa \phi^{a\ad}} F^{-2m} = 0\ .
\end{equation}
This is yet another kind of analyticity condition (S-analyticity), this time with respect to the scalars (which in fact are the coordinates on the curved manifold  $SO(6,n)/SO(6)\times SO(n)$, see Section \ref{coset}).
The second one restricts the harmonic dependence
\begin{equation}\label{secocon}
    D_{+a}{}^{-\bd} F^{-2m} = \ep_{ab} \phi^{b\bd}\frac{\pa}{\pa \phi^{++}} F^{-2m}\ .
\end{equation}
Note that if the right-hand side in (\ref{secocon}) vanished, this would be a condition defining a lowest-weight state of $SU(4)$ (or an $SO(6)$ tensor of rank $m$). However, the dependence on the scalars makes the harmonic structure in (\ref{23}) reducible.

{}From (\ref{23}) we have to extract the  irreducible harmonic structure
$\bu_{++}^{i_1 j_1} \cdots \bu_{++}^{i_{m} j_{m}}$ needed to match the conjugate
structure in (\ref{22}). It is obtained by contracting all the $u^{++}$ in (\ref{23}) with
a subset of the $\bu_{++}$, using $\bu_{++}^{ij}\; u^{++}_{kl} = -1/3\,
\delta^{[i}_k\delta^{j]}_l + \ldots$ (see (\ref{12'})). This confirms that the omitted
terms in the harmonic expansion of $\xi$ in (\ref{23}) cannot contribute - they contain higher-rank $SO(6)$ tensors. The result is
the {\it relevant part} of the function $F$, or the {\it reduced} function
\begin{equation}\label{24}
    {\cal F}^{-2m} = \sum_n   \xi_{(i_1 \cdots i_{m+n}) (j_1 \cdots j_{m+n})}\  \bu_{++}^{i_1 j_1}
    \cdots \bu_{++}^{i_{m} j_{m}}  \ \phi^{i_{m+1}j_{m+1}} \cdots \phi^{i_{m+n} j_{m+n}}\ .
\end{equation}
Notice the full symmetrization of the $i$ and $j$ indices of the $\xi$ tensor inherited
from (\ref{23}). As required, the reduced function is manifestly H-analytic (i.e., $SU(4)$ irreducible),
\begin{equation}\label{mansu4}
    D_{+a}{}^{-\bd} {\cal F}^{-2m} =  0\ .
\end{equation}
However, now the manifest S-analyticity (i.e., the dependence only on $\phi^{++}$) of (\ref{23}) is lost.

It should be made clear that (\ref{24}) is just a rearrangement of the harmonic expansion of the gauge-fixed function $F^{-2m}$. The information contained in this function is encoded in the fact that the coefficients $\xi_{(i_1 \cdots i_{m+n}) (j_1 \cdots j_{m+n})}$, which are the same in (\ref{23}) and (\ref{24}), form the $SU(4)$ irrep
$(0,m+n,0)$ \footnote{$SU(4)$ irreps with Dynkin labels $(0p0)$ are equivalent to rank $p$  symmetric traceless tensor of $SO(6)$.}. This information can be translated into two types of differential constraints on the function ${\cal F}$. In
general, the harmonic and scalar factors in (\ref{24}) form the reducible representation
$(0m0)\, \otimes\, \prod_{p=1}^n\otimes\, (010)\ \rightarrow \ (0,m+n,0) + \ldots$. The
relevant projection $(0,m+n,0)$ is obtained by symmetrizing all the $i$ and separately
all the $j$ indices (the antisymmetry of the $i$'s with the $j$'s is automatic).  Any
other irrep in this tensor product will have a subset of the $i$'s (and of the $j$'s)
antisymmetrized. The product of two $u$'s is irreducible, $(010)\otimes(010) \
\rightarrow \ (020)$ as follows form the commuting nature of the $SU(4)$ harmonics
$\bu^i_{+a}$.  The antisymmetrization of indices carried by the $\bu$'s and the $\phi$'s
is ruled out by the so-called {\it harmonicity} condition:
\begin{equation}\label{25}
    \ep^{pqrs}\frac{\pa}{\pa \bu^q_{+a}}\ \frac{\pa}{\pa \phi^{rs}_A} {\cal F} =  0\, ,
\end{equation}
where we have restored the $SO(n)$ index $A$ and suppressed the $U(1)$ charge superscript ${-2m}$. The above equation  forbids the decomposition
$(100)\otimes(010) \ \rightarrow \ (001)$. This constraint involves partial derivatives
with respect to $\bu_{+}$. Strictly speaking, such an operation is illegal in the
harmonic formalism, since the variables $u$ are not independent, as can be seen from
(\ref{12'}), (\ref{12}). However the above equation can be rewritten using covariant harmonic derivatives
introduced in (\ref{subhd}) and (\ref{cosethd}) as
\begin{equation}\label{25prime}
    \ep^{pqrs}\left(u^{+b}_q D_{+b}{}^{+a}+ \frac{1}{2} u^{+a}_q D_0 +\bu^{-\dot{b}}_q D_{-\dot{b}}{}^{+a}\right)
\ \frac{\pa}{\pa \phi^{rs}_A} {\cal F} =  0\ .
\end{equation}
Indeed, it is easy to see that this equation reduces to (\ref{25}) since our function ${\cal F}$ explicitly involves only $\bu_{+}$ harmonics. The $D_0$ term in (\ref{25prime}) is just to remove the contribution from the trace parts in $D_{+b}{}^{+a}$ as defined in (\ref{subhd}) which measures the total $U(1)$ charge $-2m$ of ${\cal F}$.  In the following however we will continue to write
the formula using
partial derivatives with respect to $\bu_{+}$.

Further, the  antisymmetrization of indices carried by the $\phi$'s is ruled out by the
constraint
\begin{equation}\label{25'}
   \ep^{pqrs} \frac{\pa}{\pa \phi^{pq}_A} \ \frac{\pa}{\pa \phi^{rt}_B}\ {\cal F} =  0
\end{equation}
which forbids the decomposition $(010)\otimes(010) \ \rightarrow \ (101)\oplus (000)$.
In $SO(6)$ (vector) notation (\ref{25'}) reads
\begin{eqnarray}
  &&  \frac{\pa}{\pa \phi^{[M}_A} \ \frac{\pa}{\pa \phi^{N]}_B}\ {\cal F} =   0\ , \label{trivial}\\
  &&  \frac{\pa}{\pa \phi^{M}_A} \ \frac{\pa}{\pa \phi^{M}_B}\ {\cal F} =
    0\ . \label{25''}
\end{eqnarray}
Here we do not take into account the fact that the physical scalars $\phi$ parametrize a
curved manifold and hence the derivatives in (\ref{25'}) should be considered covariant
with respect to the metric of the manifold. In Section \ref{coset} we show that this
leads to a modification of (\ref{25''}) by a term proportional to
$\delta_{AB}$.

\section{The coset of physical scalars}\label{coset}

\subsection{The coset $SO(6,n)/SO(6)\times SO(n)$}\label{subcoset}

Here we  briefly recall why the scalars of ${\cal N}=4$ Poincar\'{e} supergravity
describe the coset space $SO(6,n)/SO(6)\times SO(n)$ \cite{Cremmer:1977tt,Bergshoeff:1980is,de Roo:1984gd}.

${\cal N}=4$ Poincar\'{e} supergravity is obtained by coupling the {\it off-shell} Weyl
multiplet to $6+n$ free vector multiplets. The first six are compensating multiplets
(i.e. their kinetic terms have the wrong sign), the remaining $n$ are physical. Each
vector multiplet supplies 6 scalars, so the total number is $6(6+n)$. We denote the
first $6\times 6$ by $\varphi^M_N$ and the other $6\times n$ by $\phi^M_A$, where $M,N =
1\ldots 6$ and $A=1\ldots n$ are $SO(6)$ and $SO(n)$ vector indices, respectively.

The Weyl multiplet contains an auxiliary field $D_{MN} = D_{NM}$, $D_{MM}=0$ in the
\underline{\bf 20'} of $SO(6) \sim SU(4)$. It serves as a Lagrange multiplier for the
following quadratic combination of scalars:
\begin{equation}\label{26}
    D_{MN}\ (\varphi^M_K \varphi^N_K - \phi^M_A \phi^N_A)\ .
\end{equation}
So, it imposes an algebraic constraint which eliminates 20 of the scalars. In addition,
one makes a Weyl (dilatation) gauge choice for the trace of the quadratic form in
(\ref{26}), thus fixing yet another scalar. So, the resulting condition is (up to
normalization)
\begin{equation}\label{27}
    \varphi^M_K \varphi^N_K - \phi^M_A \phi^N_A = \delta^{MN}\ .
\end{equation}
Notice that this condition is invariant under {\it local} $SO(6)$ which allows to gauge
away 15 additional scalars. Altogether 36 scalars are eliminated and the remaining $6n$
do indeed parametrize the coset $SO(6,n)/SO(6)\times SO(n)$.

Conditions (\ref{27}) can be solved by first fixing an $SO(6)$ gauge such that the
$6\times 6$ matrix $\varphi^M_N$ becomes symmetric, $\varphi = \varphi^T$, after which
one can write down
\begin{equation}\label{28}
    \varphi = \sqrt{\mathbb{I} + \phi \phi^T}\ .
\end{equation}
We can say that the $6n$ physical scalars $\phi$ are the {\it unconstrained} coordinates
on  the coset $SO(6,n)/SO(6)\times SO(n)$.

\subsection{Harmonic description}

The higher-derivative terms (\ref{19}), (\ref{19'}) involve the function (potential) $F$
defined on the coset of physical scalars. The peculiarity of this function is that it
depends only on a single projection $Y^{++}_A = \phi^{++}_A(x,u) + \ldots$ of the six-vectors of coset coordinates,
obtained with the help of the $SU(4)$ harmonic variables. This is a typical example of
an {\it analytic harmonic realization} of a coset space. Another, very similar example
is that of the $\cN=4$ superconformal group $PSU(2,2/4)$ realized on the Grassmann
analytic superfields (\ref{01}) (see Section \ref{css}). Here we explain this coset construction, following closely the case of $\cN=2$ superconformal symmetry and Poincar\'e supergravity \cite{Galperin:1985zv,Galperin:1987ek,Galperin:2001uw} and of $\cN=2$ quaternionic sigma models \cite{Bagger:1987rc,Galperin:1992pj,Ivanov:1999vg}.

We start by writing down the algebra of $SO(6,n)$ in a basis suitable for the
forthcoming introduction of the harmonic variables (\ref{12'}). The $SO(n)$ generators
are $M_{AB}=-M_{BA}$ and the $SO(6)$ ones are written in an $S(U(2)\times U(2))$ basis. Thus, the $S(U(2)\times U(2))$ generators are
\begin{equation}\label{29}
    Z_{+a}{}^{+b}\ (Z_{+a}{}^{+a}=0)\,, \quad Z_{-\ad}{}^{-\bd}\ (Z_{-\ad}{}^{-\ad}=0)\,, \quad Z_0 \ ,
\end{equation}
and the remaining generators of $SU(4)$ are $Z_{+a}{}^{-\bd}$ and
$Z_{-\ad}{}^{+b}$. Finally,  the generators of the coset $SO(6,n)/SO(6)\times SO(n)$ are
$L_{A\, a\ad}$, $L_{A\, ++}$ and $L_{A\, --}$. Then the algebra of $SO(6,n)$ takes the
form
{\allowdisplaybreaks
\begin{eqnarray}
  [L_{A\, a\ad},L_{B\, b\bd}] &=& \delta_{AB}\left(\ep_{\ad\bd} \ep_{(ac} Z_{+b)}{}^{+c} + \ep_{ab} \ep_{(\ad \dot c} Z_{-\dot b)}{}^{-\dot c}\right) + \ep_{ab}\ep_{\ad\bd} M_{AB} \nn\\
  {[}L_{A\, a\ad},L_{B\, ++}{]} &=&  \delta_{AB} \ep_{\ad\bd}Z_{+a}{}^{-\bd}\nn\\
  {[}L_{A\, a\ad},L_{B\, --}{]} &=&  \delta_{AB}\ep_{ab} Z_{-\ad}{}^{+b} \nn\\
  {[}L_{A\, ++},L_{B\, --}{]} &=&  \delta_{AB}Z_0 + M_{AB} \nn\\
  {[}Z_{+a}{}^{+b} , L_{A\, c\dot c}{]} &=& \delta^b_c L_{A\, a\dot c} - \frac{1}{2} \delta^b_a L_{A\, c\dot c}\nn\\
  {[}Z_{-\ad}{}^{-\bd} , L_{A\, c\dot c}{]} &=& \delta^{\bd}_{\dot c} L_{A\, c\dot a} - \frac{1}{2} \delta^{\bd}_{\dot a} L_{A\, c\dot c} \nn\\
  {[}Z_0 , L_{A\, \pm\pm}{]} &=& \pm 2 L_{A\, \pm\pm} \label{algebra}\\
  {[}Z_{+a}{}^{-\bd}, L_{A\, --}{]} &=& \ep^{\bd \dot c} L_{A\, a\dot c} \nn\\
  {[}Z_{-\ad}{}^{+b}, L_{A\, ++}{]} &=& \ep^{b c} L_{A\, c\dot a} \nn\\
  {[}Z_{+a}{}^{-\bd}, L_{A\, c\dot c}{]} &=& \ep_{ac} \delta^{\bd}_{\dot a} L_{A\, ++} \nn\\
  {[}Z_{-\ad}{}^{+b}, L_{A\, c\dot c}{]} &=& \ep_{\ad\dot c} \delta^{b}_{c} L_{A\, --}\nn\\
  {[}Z_{+a}{}^{+b} , Z_{+c}{}^{+d} {]} &=& \delta^b_c Z_{+a}{}^{+d} - \delta^d_a Z_{+c}{}^{+b} \nn\\
  {[}Z_{-\ad}{}^{-\bd} , Z_{-\dot c}{}^{-\dot d} {]} &=& \delta^{\bd}_{\dot c} Z_{-\dot a}{}^{-\dot d} - \delta_{\ad}^{\dot d} Z_{\dot c}{}^{-\dot b}  \nn\\
  {[}Z_0 , Z_{+a}{}^{+b} {]} &=& 2 Z_{+a}{}^{+b} \nn\\
  {[}Z_0 , Z_{-\ad}{}^{+b} {]} &=& -2 Z_{-\ad}{}^{+b} \nn
\end{eqnarray}
}
Now, we want to realize this algebra on a coset of the group $SO(6,n)$. The standard
coset $SO(6,n)/SO(6) \times SO(n)$ is obtained by putting all the generators $M$ and $Z$
in the coset denominator and leaving all the $L$'s in the coset with associated $6n$
coordinates $\phi$:
\begin{equation}\label{30'}
     \frac{SO(6,n)}{(M,Z)}\ \sim\ \{ \phi^{++}_A\,,\phi^{--}_A\,, \phi^{a\ad}_A\}\ .
\end{equation}

We wish to have an alternative {\it S-analytic} coset involving only the coordinates $\phi^{++}_A$
associated with the generators $L_{A\, ++}$. To this end we have to move the generators
$L_{A\, a\ad}$, $L_{A\, --}$ to the coset denominator. In doing this we encounter a
problem: The $SO(6)$ generator $Z_{+a}{}^{-\bd}$ converts $L_{A\, a\ad}$ into the coset
generator $L_{A\, ++}$. In order to avoid this, we proceed to the `harmonization' of the
coset. This means to introduce an additional group $\widehat{SU(4)}$ which we treat as
independent of the $SO(6)$ from the coset denominator. Let us denote its generators by
$T_{+a}{}^{+b}$, $T_{-\ad}{}^{-\bd}$, $T_0$, $T_{+a}{}^{-\bd}$ and $T_{-\ad}{}^{+b}$, in
complete analogy with $SO(6)$. We assume that this extra $\widehat{SU(4)}$ acts as an
external automorphism of  (\ref{algebra}), i.e. $[T,Z] = Z$, $[T,L] = L$. Then it is
clear that the combination $Z_{+a}{}^{-\bd} - T_{+a}{}^{-\bd}$ commutes with the
generators of (\ref{algebra}), in particular, with  $L_{A\, a\ad}$. So, to avoid the
above problem, we replace $Z_{+a}{}^{-\bd}$ in the coset denominator by this
combination. The group $\widehat{SU(4)}$ is itself realized on the harmonic coset
$\widehat{SU(4)}/S(\widehat{U(2)}\times \widehat{U(2)})$, which means that we have to add the generators of
the automorphism subgroup $S(\widehat{U(2)}\times \widehat{U(2)})$ to  the coset denominator. The
result is a particular {\it S-analytic} realization of the coset
\begin{equation}\label{30}
    \frac{SO(6,n)\subset\hskip-11pt\times\ \widehat{SU(4)}}{(M,L_{a\ad}, L_{--},  Z_+{}^+, Z_-{}^-, Z_0, Z_-{}^+,
    Z_+{}^--T_+{}^-, T_+{}^+, T_-{}^-, T_0)}\ \sim (\phi^{++}_A, w_i^{+a}, w_i^{-\ad})
\end{equation}
parametrized by the coordinates $\phi^{++}_A$ associated with the $SO(6,n)$  generators
$L_{A\, ++}$ and by harmonics $w_i^{+a}, w_i^{-\ad}$ (the latter differ from the usual
${SU(4)}$ harmonics $u$ (\ref{12'}), as explained below).

This coset is analytic in the sense that we consider functions $F(\phi^{++}_A,w)$  on it
which are annihilated by the generators $L_{A\, a\ad}$ and $L_{A\, --}$. Then the
algebra  (\ref{algebra}) implies
\begin{equation}\label{31}
    L_{A\, a\ad}F = L_{A\, --} F = 0 \ \ \Rightarrow \ \ M_{AB}F = Z_{+a}{}^{+b}F = Z_{-\ad}{}^{-\bd} F = Z_{-\ad}{}^{+b} F =0\,,
\end{equation}
i.e., $F$ cannot carry $SO(n)\times SU(2)\times SU(2)$ indices, but can have $U(1)$
charges under both $Z_0$ and $T_0$. In addition, we impose the coset defining constraint
\begin{equation}\label{32}
    (Z_{+a}{}^{-\bd} - T_{+a}{}^{-\bd}) F = 0 \ .
\end{equation}
It leads to a particular mixing of the coordinates associated with
the $SO(6)$ generators $Z$ and with the $\widehat{SU(4)}$ generators $T$. For this
reason (\ref{30}) is a semi-direct product (denoted by $\subset\hskip-11pt\times$ in (\ref{30})) of the two cosets  $SO(6,n)/SO(6) \times
SO(n)$ and   $\widehat{SU(4)}/S(\widehat{U(2)}\times \widehat{U(2)})$.

The actual construction of the coset goes through the following steps. We first
introduce a {\it double harmonic space} involving,  in addition to the $\widehat{SU(4)}$
harmonic variables $u$, harmonics $\kappa_I{}^i$ on ${SU(4)} \sim SO(6)$ satisfying the
defining conditions (cf. (\ref{12'}))
\begin{equation}\label{33}
    \kappa_I{}^i \bar\kappa_i{}^J = \delta^J_I\,, \quad \bar\kappa_i{}^I \kappa_I{}^j = \delta^j_i\,, \quad \ep^{IJKL}\kappa_I{}^i\kappa_J{}^j\kappa_K{}^k\kappa_L{}^l = \ep^{ijkl}\ .
\end{equation}
They undergo $SU(4)$ transformations of two types: local (in the sense of ${SU(4)} \sim
SO(6)$ from the coset denominator)  with parameter $\lambda$ and rigid with parameter
$\sigma$:
\begin{equation}\label{34}
    \delta \kappa_I{}^i = \lambda_I{}^J \kappa_J{}^i + \kappa_I{}^j \sigma_j{}^i \ .
\end{equation}

Our task now will be to make a change of variables from $\kappa, u$ to $z, w$ which are
inert under the rigid $SU(4)$ and have simple transformation properties under the local
$SU(4)$. This will allow us to impose the coset constraint (\ref{32}) in a covariant
way. We start by projecting the harmonics $\kappa$ with $u, \bu$:
\begin{eqnarray}
  && \kappa_{+a}{}^{+b} = \bu_{+a}{}^{I} \kappa_I{}^i u_i{}^{+b}\,, \quad  \kappa_{+a}{}^{-\bd} = \bu_{+a}{}^{I} \kappa_I{}^i u_i{}^{-\bd} \nn\\
  && \kappa_{-\ad}{}^{+b} = \bu_{-\ad}{}^{I} \kappa_I{}^i u_i{}^{+b}\,, \quad  \kappa_{-\ad}{}^{-\bd} = \bu_{-\ad}{}^{I} \kappa_I{}^i u_i{}^{-\bd}
\end{eqnarray}
and similarly for the conjugate matrix $\bar\kappa$. Next we make the following non-linear
change of variables (to simplify the notation, we suppress the $SU(2)\times SU(2)$
indices; the position of the $U(1)$ charges allows to unambiguously restore them):
\begin{eqnarray}
  && z_-{}^+ = \kappa_-{}^+ (\kappa_+{}^+)^{-1} = - (\bar\kappa_-{}^-)^{-1}\bar\kappa_-{}^+\,, \quad z_-{}^- = \bar\kappa_-{}^-  \nn \\
  && z_+{}^- = \kappa_+{}^- (\kappa_-{}^-)^{-1} = - (\kappa_+{}^+)^{-1}\bar\kappa_+{}^-\,, \quad z_+{}^+ = \bar\kappa_+{}^+\ .
\end{eqnarray}
These new variables satisfy an algebraic constraint following from the fact that $\kappa \in
SU(4)$, i.e. $\det \kappa =1$. It can be used to eliminate, e.g.  $\det z_-{}^-$ while the
remaining  $z_0 \equiv \det z_+{}^+$ can be treated as the coordinate of the $U(1)$
factor in $S(U(2)\times U(2)) \subset SU(4)$.

It is then not hard to check that the new variables $z$ transform in the following way
under the local $SU(4)$:
\begin{eqnarray}
  && \delta z_-{}^+ = \hat\lambda_-{}^+ \,, \quad \delta z_-{}^- = z_-{}^-\hat\lambda_-{}^-   \,, \quad \delta z_+{}^+ = \hat\lambda_+{}^+z_+{}^+ \nn \\
  && \delta z_+{}^- = \hat\lambda_+{}^+z_+{}^- + z_+{}^-\hat\lambda_-{}^- - \hat\lambda_+{}^-\,, \label{deltaz}
\end{eqnarray}
where $\hat\lambda_\pm{}^\pm = \bar w_\pm{}^I \lambda_I{}^J w_J{}^\pm$ and we have
introduced the {\it new harmonics}
\begin{eqnarray}
  && w_i{}^{+a} = u_i{}^{+a} + u_i{}^{-\bd}z_{-\bd}{}^{+a}\,, \quad  w_i{}^{-\ad} = u_i{}^{-\ad} \nn\\
  && \bar w_{+a}{}^i = \bar u_{+a}{}^i\,, \quad  \bar w_{-\ad}{}^i = \bar u_{-\ad}{}^i - z_{-\ad}{}^{+b} \bu_{+b}{}^i \label{wharm}
\end{eqnarray}
with transformation laws
\begin{eqnarray}
  && \delta w_i{}^{+a} = w_i{}^{-\bd}\hat\lambda_{-\bd}{}^{+a}\,, \quad  \delta w_i{}^{-\ad} = 0\nn\\
  && \delta \bar w_{+a}{}^i = 0\,, \quad  \delta \bar w_{-\ad}{}^i = -\hat\lambda_{-\ad}{}^{+b} \bar w_{+b}{}^i \ . \label{trw}
\end{eqnarray}
We point out that these new harmonics are not unitary anymore (i.e., $\bar w$ is not the
conjugate of $w$), but they still satisfy the same algebraic relations as the unitary
harmonics $u$ (\ref{12'}).

What we have achieved is that the new variables do not mix under the local $SU(4)$
transformations with parameters $\hat\lambda$. This allows us to eliminate  all of the
$z$ variables (with the exception of $z_0$) in a covariant way, which corresponds to
imposing the $Z$ coset conditions from (\ref{31}) and the $Z-T$ condition (\ref{32}).

\subsection{Covariant constraints on the function ${F}$}

Now we are able to see how the naive constraints (\ref{25}), (\ref{trivial}), (\ref{25''}) are modified due to
the curvature of the coset space (\ref{30}) on which the reduced function ${\cal F}$ (\ref{24}) lives. The origin of these constraints can be traced back to the S-analyticity conditions satisfied by the gauge-fixed function $F$ (\ref{23}). On the curved manifold they become {covariant} constraints (cf. (\ref{31})):
\begin{equation}\label{covsancon}
    {\cal D}_{A\, a\ad} F = {\cal D}_{A\, --} F = 0\ .
\end{equation}
Here ${\cal D}_{A\, M}$ are covariant derivatives generalizing the flat derivatives
$\pa/\pa \phi$. They satisfy the same $SO(6,n)$ algebra as the generators $L_{A\, M}$.

Let us start with the constraint  (\ref{25''}). The second-order derivative in it can be rewritten as follows:
\begin{eqnarray}
  {\cal D}_{A\, M}{\cal D}_{B\, M} F &=& ({\cal D}_{A++} {\cal D}_{B--} + {\cal D}_{A--} {\cal D}_{B++} -2 {\cal D}_{A a\ad}\ep^{ab}\ep^{\ad\bd}{\cal D}_{Bb\bd}) F \nn\\
  &=& [{\cal D}_{A--} , {\cal D}_{B++}]  F = -\delta_{AB} Z_0 F \,, \label{correctcons}
\end{eqnarray}
where we have used the S-analyticity constraints (\ref{covsancon}) and the algebra (\ref{algebra}). The function $F^{-4(g-1)}$ has two independent $U(1)$ charges, one with respect to the generator $T_0$, $T_0 F^{-4(g-1)} = -4(g-1) F^{-4(g-1)}$ and the
other for $Z_0$. For a reason which will become clear in the next section, the $Z_0$ charge takes a different value, $Z_0 F = -4(g+1) F$. Thus, we have
\begin{equation}\label{modconstr}
    {\cal D}_{A\, M}{\cal D}_{B\, M} F = 4(g+1)\delta_{AB} F\ ,
\end{equation}
or, in $SU(4)$ notation,
\begin{equation}\label{conts}
   \ep^{pqrs} {\cal D}_{Apq} {\cal D}_{Brs}\ {F} =  32(g+1)\delta_{AB}\,{F}\ .
\end{equation}

Further, the second-order derivative in
(\ref{trivial}) is replaced by ${\cal D}_{A\, [M}{\cal D}_{B\, N]} {F}$. Due to the constraints (\ref{covsancon}), this operator has only two non-vanishing projections obtained by taking $M=++$ and $N=--$ or
$N=a\ad$. The first choice yields back the constraint (\ref{correctcons}), while the
second gives rise to the commutator
\begin{equation}\label{commu}
    [{\cal D}_{Aa\ad} , {\cal D}_{B++}] {F} =  \delta_{AB}\, \ep_{\ad\bd} {\cal Z}_{+a}{}^{-\bd} {F}
\end{equation}
where ${\cal Z}$ is the covariant derivative replacing the generator $Z$. The effect of this is just a particular $SO(6)$ transformation of the coset coordinates, hence it is not really a constraint on the function.

Finally, in eq. (\ref{25}) (or (\ref{25prime})) the flat partial derivative with respect to scalars is replaced by a covariant derivative
\begin{equation}
    \ep^{pqrs}\frac{\pa}{\pa \bu^q_{+a}}\ {\cal D}_{Ars} {\cal F} =  0\,.\label{covfirstorderder}
\end{equation}

We would like to point out that in the string theory analysis given in the following sections, the
differential equations are obtained on functions ${\cal F}$  which is the relevant part of $F$ that survives the harmonic space integrals.  Indeed string theory amplitudes directly see ${\cal F}$.
The crucial step used in equation (\ref{correctcons}) was that $F$ does not depend on 5 combinations of moduli as is expressed in the S-analyticity constraint (\ref{covsancon}). It is easy to see
that ${\cal F}$ does not satisfy this S-analyticity constraint since it is obtained by making a
certain $SU(4)$ projection on $F$.  Therefore the individual steps in this derivation cannot be applied to ${\cal F}$. However, the second order differential operators considered here are
not sensitive to any particular $SU(4)$ projection of $F$ and therefore the final equations are
still true on ${\cal F}$.

\section{$\cN=4$ conformal supersymmetry and supergravity}\label{css}

Here we show that the realization of G-analytic superfields of the type (\ref{01}) as
functions on a particular coset of the $\cN=4$ conformal superalgebra $PSU(2,2/4)$ is
very similar to the bosonic coset construction of the preceding section. This algebra
involves the generators of Lorentz transformations ($M_{\m\n}$), translations ($P_\m$),
conformal boosts ($K^\m$), dilatation ($D$), R symmetry $SU(4)$ ($Z_i^j$), Poincar\'{e}
supersymmetry ($Q^\a_i$ and $\bar Q_\da^i$) and special conformal supersymmetry
($S_\a^i$ and $\bar S^\da_i$) with anticommutation relations for the odd generators
(schematically)
\begin{eqnarray}
  \{Q, \bar Q\} &=& P \nn\\
  \{S, \bar S\} &=& K \nn\\
  \{Q, S\} &=& M +D + Z \ .
\end{eqnarray}
The standard superspace corresponds to the coset
\begin{equation}\label{44}
    \frac{PSU(2,2/4)}{(M,K,D,S, \bar S, Z)}\ \sim (x^\m, \q_\a^i, \bq^\da_i)
\end{equation}
involving all the 16 Grassmann variables associated with the supersymmetry generators.
In order to obtain G-analytic superfields depending on half of these Grassmann variables, we add the  $SU(4)$ harmonic projections of
the $Q$ generators $Q^\a_{-\ad} = \bu^i_{-\ad}\, Q^\a_i $ and $\bar Q_\da^{+a} = \bar
Q_\da^i\, u^{+a}_i $ to the coset denominator, thus leaving only the odd coordinates
$\q_\a^{+a}$ and $\bq^\da_{-\ad}$ in the coset. However, exactly as in the bosonic case
of Section \ref{coset}, the $SU(4)$ generator $Z_+{}^-$ converts $Q_-$ and $\bar Q^+$
from the coset denominator into the coset generators $Q_+$ and $\bar Q^-$. In order to
avoid this, we introduce the external automorphism group $\widehat{SU(4)}$ with
generators $T$. Then the combination $Z_+{}^--T_+{}^-$ commutes with all the $Q$'s and
thus can be safely put in the coset denominator:\footnote{Here we follow the formulation of $\cN=2$ conformal supersymmetry of \cite{Galperin:1985zv,Galperin:2001uw}. A somewhat different approach is proposed in \cite{Hartwell:1994rp}.}
\begin{equation}\label{45}
    \frac{PSU(2,2/4)\subset\hskip-11pt\times\  \widehat{SU(4)}}{(M,K,D,S, \bar S, Q_-, \bar Q^+,  Z_+{}^+, Z_-{}^-, Z_0, Z_-{}^+, Z_+{}^--T_+{}^-, T_+{}^+, T_-{}^-, T_0)}\ \sim (x, \q^+, \bq_-, w)\ .
\end{equation}
Here the harmonics $w$ are defined in exactly the same way as in Section \ref{coset},
eq.~(\ref{wharm}), replacing the $SO(6)$ harmonics $\kappa$ by R-symmetry $SU(4)$ harmonics. They transform as in (\ref{trw}) with the parameter $\hat\lambda$
replaced by the G-analytic superparameter
\begin{equation}\label{46}
    \Lambda_{-\bd}{}^{+a}(x, \q^+, \bq_-, w) = \bar w_{-\bd}{}^i \lambda_i{}^j w_j{}^{+a} + i\q^{+a}\sigma^\m \bq_{-\bd} k_\m + i   \bar w_{-\bd}{}^i \eta^\a_i \q^{+a}_\a + i \bq_{-\bd}^\da \bar\eta_\da^i w_i{}^{+a}
\end{equation}
containing the parameters $\lambda$ of the R-symmetry $SU(4)$, $k$ of conformal boosts and $\eta$ of special conformal supersymmetry.

The basic G-analytic conformal superfield $Y^{++}(x, \q^+, \bq_-, w)$ (\ref{01}) (with
superconformal harmonics $w$ instead of $u$) describes the vector supermultiplet. It
transforms with a G-analytic superconformal weight factor:
\begin{eqnarray}
  &&  \delta Y^{++} = {Y^{++}}'(x',\q',\bq',w') - Y^{++}(x,\q,\bq,w) = \Lambda Y^{++}   \label{47}\\
  &&  \Lambda(x, \q^+, \bq_-, w) = \rho +k_\m x^\m + \bar w_{+a}{}^i \lambda_i{}^j w_j{}^{+a} + i   \bar w_{+a}{}^i \eta^\a_i \q^{+a}_\a  + i \bq_{-\bd}^\da \bar\eta_\da^i w_i{}^{+a} \nn
\end{eqnarray}
where $\rho$ is the parameter of dilatations.\footnote{It can be shown that $\Lambda_{-\bd}{}^{+a} = D_{-\bd}{}^{+a}\Lambda$.} The other G-analytic object we are
discussing here is the descendant $K^{++}_{\m\n}$ (\ref{16}) of the Weyl multiplet. It
is superconformal covariant due to the on-shell constraint and transforms with weight
two, according to its scaling dimension, $\delta K^{++} = 2 \Lambda K^{++}$.

The generalization to $\cN=4$ conformal supergravity is done by replacing the parameters
$\Lambda_{-\bd}{}^{+a}$ and $\Lambda$ by arbitrary G-analytic superfields. Poincar\'{e}
supergravity is obtained by coupling the Weyl multiplet to a set of six compensating
vector multiplets (cf. (\ref{01}))
\begin{equation}\label{48}
     y^{++}_{ij}(x, \q^+, \bq_-, w) = - y^{++}_{ji} = \varphi^{kl}_{ij} w_k^{+a} w_l^{+b} \ep_{ab} + \mbox{$\q$ terms}\ .
\end{equation}
Here we see the $6\times 6$ matrix of compensating scalars $\varphi^{kl}_{ij}$. Let us
consider the following projections of $y^{++}_{ij}$ with the harmonics $w$:
\begin{equation}\label{49}
    y^{++}_{a\ad} = \bar w^i_{+a} \bar w^j_{-\ad}\ y^{++}_{ij}\,, \qquad y_0 = \half \ep^{ab} \bar w^i_{+a}\bar w^j_{+b}\ y^{++}_{ij}\ .
\end{equation}
It is easy to check that they transform as follows:
\begin{equation}\label{49'}
    \delta y^{++}_{a\ad} = \ep_{ab} \Lambda_{-\ad}{}^{+b} \, y_0 + \Lambda\, y^{++}_{a\ad}\,, \qquad \delta y_0 = \Lambda\, y_0\ ,
\end{equation}
so their ratio transforms as a {\it compensator} for the local superconformal
transformations:
\begin{equation}\label{50}
    \delta \left(\frac{y^{++}_{a\ad}}{y_0} \right) = \ep_{ab} \Lambda_{-\ad}{}^{+b}\ .
\end{equation}
Then, with the help of this compensator we can define new harmonics {\it inert under the
local superconformal transformations} (notice the similarity with (\ref{wharm}) and (\ref{trw})):
\begin{eqnarray}
  && v_i{}^{+a} = w_i{}^{+a} - w_i{}^{-\bd}\ep^{ab}\frac{y^{++}_{b\bd}}{y_0}\,, \quad  v_i{}^{-\ad} = w_i{}^{-\ad} \nn\\
  && \bar v_{+a}{}^i = \bar w_{+a}{}^i\,, \quad  \bar v_{-\ad}{}^i = \bar w_{-\ad}{}^i + \ep^{ab}\frac{y^{++}_{b\ad}}{y_0} \bar w_{+a}{}^i \label{vharm}\\
  && \delta v = \delta \bar v = 0 \ . \nn
\end{eqnarray}

The role of the compensators is to completely absorb the local superconformal transformations. This allows us to use the parameter $\Lambda_{-\ad}{}^{+b}$ in (\ref{50}) fix a gauge in which $y^{++}_{a\ad}=0$, thus identifying the harmonics $v$ and $w$. This means, in particular, that the conformal $SU(4)$ (generators $Z$ in (\ref{45})) is identified with $\widehat{SU(4)}$ (generators $T$ in (\ref{45})). By the same logic, we can use the parameter $\hat\lambda_-{}^+ $ of local $SO(6)$ transformations in (\ref{deltaz}) to gauge away the compensator $z_-{}^+$. This results in the identification of the harmonics $w$ with $u$. So, at the expense of manifest covariance, the different $SU(4)$ groups discussed above are reduced to a unique one, and the harmonics to the original ones (\ref{12'}). This gauge fixing procedure establishes a bridge between the S-analytic coset (\ref{30}) and the G-analytic coset (\ref{45}).

Finally, we are ready for the superconformal covariantization of the higher-derivative
terms (\ref{19}), (\ref{19'}). It is achieved in three steps. Firstly, we replace the explicit
harmonics $u$ in $F(Y,u)$ by the new inert ones $v$ (however, the superfields $Y$ still
depend on the conformal harmonics $w$). Secondly, we introduce weightless G-analytic
superfields $Y/y_0$. In this way the potential $F(Y,v)$ becomes conformal invariant.
Thirdly, we use the G-analytic density $y_0$ to compensate the weight $4(g+1)$ of the
Weyl factor (the measure is weightless, as can be seen from its vanishing scaling dimension).
The result is
\begin{eqnarray}
 S_1 &=& \int \, d^4x \, du \, d^4\q^+ d^4\bq_{-} \, (\bar K^{++}_{\m\n} \bar K^{++\, \m\n}) \, (y_0)^{-4(g+1)} \,   (K^{++}_{\rho\sigma} K^{++\, \rho\sigma})^{g}\, F_1^{-4(g-1)}\left(\frac{Y^{++}_A}{y_0},v\right)\,,\nn\\
  S_2 &=& \int \, d^4x \, du \, d^4\q^+ d^4\bq_{-} \, (K^{++}_{\m\n} K^{++\, \m\n})^{g+1} \, (y_0)^{-4(g+1)} \,   F_2^{-4(g-1)}\left(\frac{Y^{++}_A}{y_0},v\right)\ . \label{19bis}
\end{eqnarray}

The presence of the density $(y_0)^{-4(g+1)}$ in (\ref{19bis}) explains why in (\ref{modconstr}) we took the value  $Z_0 F = -4(g+1) F$ of the charge $Z_0$, different from that of the charge $T_0$. This density should be viewed as part of the covariantized function $F$ discussed at the end of Section \ref{coset}. Then, $F$ is a function of the G-analytic superfields $Y^{++}_A$ and $y^{++}_{ij}$ and hence is a G-analytic superconformal object itself. This means that it is annihilated by the supercharges $Q_-, \bar Q^+$ from the coset denominator in (\ref{45}). This is compatible with the condition of superconformal primarity (that the object is annihilated by all the special superconformal charges $S$) only if the dilatation and $Z_0$ weights of the object coincide \cite{Galperin:2001uw,Ferrara:2000eb}. Finally, the local $SU(4)$ gauge-fixing procedure (elimination of the compensators) results in the identification of the $Z_0$ charges from (\ref{30}) and (\ref{45}). The automorphism charge $T_0$ remains independent and, indeed, takes a different value.\footnote{This situation is different from $\cN=2$ superconformal symmetry where the relevant G-analytic superfields, e.g. the hypermultiplet, have equal $Z_0$ and $T_0$ charges \cite{Galperin:2001uw}. This can be explained by the different properties of the G-analytic superspace measures -- the  $\cN=2$ measure has a conformal weight while the  $\cN=4$ one does not.}



\section{Topological amplitudes - review}\label{Sect:Review}
In Sections \ref{global} and \ref{coset} it was argued from the general structure (\ref{24}) of the harmonic expansion of the supergravity amplitudes $\mathcal{F}^{(1,3)}_g$ that they fulfill differential equations of first order (\ref{25}) and second order (\ref{conts}) in the moduli of the internal compactification manifold (i.e. $K3\times T^2$ for type II string theory). In this section, we would like to check these relations by applying them directly to the string amplitudes. Since, as we have already pointed out, the latter are captured by correlation functions of the topological string, it would be logical, to consider the twisted version of the theory. However, here we are facing the problem that some of the moduli involved in the $K3\times T^2$ compactification are in fact part of the Ramond-Ramond sector of the theory, for which we have at present no representation in terms of the ${\cal N}=4$ superconformal algebra, which is used to formulate the topological correlators. Besides that, the direct study of (\ref{25}) and (\ref{conts}) in the untwisted version of the type II string is quite cumbersome, since we would have to deal with (in principle) an arbitrary high number of loops.

Fortunately, as was found in \cite{Antoniadis:2006mr}, the dual amplitudes of the couplings (\ref{19'}) in the heterotic theory compactified on $T^6$ begin receiving corrections already at the 1-loop level, which are relatively simple to compute. Therefore, for the purpose of checking (\ref{25}) and (\ref{conts}), we will focus on this amplitude which we review below.

After performing explicitly the superspace integrals of the $1/2$-BPS $F$-type term (\ref{19'}) we encounter among many different contributions a coupling of two self-dual Riemann tensors, two graviscalars and $2g-2$ graviphoton field strengths at $(g+1)$-loop order
\begin{align}
S_2=\int d^4x\mathcal{F}_g^{(3)} R_{(+)}^2(\partial\partial\Phi)^2(T_{(+)}^{++})^{2g-2}\, ,
\end{align}
where we remind that $\mathcal{F}_g^{(3)}$ corresponds in the supergravity context to the reduced part of $F_2^{-4(g-1)}$.
The corresponding heterotic string 1-loop torus amplitude can be formulated as the following two-dimensional integral over the fundamental domain $\mathbb{F}$ of the world-sheet torus
\begin{align}
\Fg=\int_{\mathbb{F}} \frac{d^2\tau}{\bar{\eta}^{24}}\tau_2^{2g-1} G_{g+1}\sum_{(P^L,P^R) \in \Gamma^{(6,22)}} \left( \frac{1}{2}\bar{u}_{++}^{ij}P^L_{ij} \right)^{2g-2}q^{\frac{1}{2}P_L^2}\bar{q}^{\frac{1}{2}P_R^2}.\label{hetamp}
\end{align}
In this expression $\tau=\tau_1+i\tau_2$ is the Teichm\"uller parameter of the torus, while $q=e^{2\pi i\tau}$. Moreover, $\eta(\tau)$ is the Dedekind eta-function given by
\begin{align}
\eta(\tau)=q^{\frac{1}{24}}\prod_{n=1}^\infty(1-q^n),
\end{align}
and $G_{g+1}$ is defined via the following expansion of a generating functional for space-time correlation functions
\begin{align}
G(\lambda,\tau,\bar{\tau})&\equiv \sum_{h=0}^\infty\frac{1}{(h!)^2}\left(\frac{\lambda}{\tau_2}\right)^{2h}
\langle\prod_{i=1}^h\int d^2x_iX^1\bar{\partial} X^2(x_i)\prod_{j=1}^{h}
\int d^2y_j\bar{X}^2\bar{\partial}\bar{X}^1(y_j)\rangle=\nonumber\\
&=\sum_{h=1}^\infty\lambda^{2h}G_{h}(\tau,\bar{\tau}).
\end{align}
In \cite{Antoniadis:1995zn}, this generating functional was calculated with the result
\begin{align}
G(\lambda,\tau,\bar{\tau})=\left(\frac{2\pi i\lambda\bar{\eta}^3}{\bar{\vartheta}
(\lambda,\bar{\tau})}\right)^2
\text{exp}\left(-\frac{\pi\lambda^2}{\tau_2}\right),
\end{align}
where $\vartheta$ is the usual odd theta-function defined by
\begin{align}
\vartheta(z,\tau)=\sum_{n\in\mathbb{Z}}q^{\frac{1}{2}\left(n-\frac{1}{2}\right)^2}e^{2\pi i
\left(z-\frac{1}{2}\right)\left(n-\frac{1}{2}\right)}.
\end{align}
The most important property of $G_g$ for our purposes is the fact that upon differentiation with respect to $\tau$ it becomes $G_{g-1}$:
\begin{align}
&\frac{\partial}{\partial \tau} G_g= -\frac{i\pi}{2\tau_2^2} G_{g-1}.\label{Ggeq}
\end{align}

In (\ref{hetamp}), $\bar{u}_{++}^{ij}$ are precisely the harmonics of the coset $\frac{SU(4)}{S(U(2)\times U(2))}$, which appear in the reduced harmonic expansion of the $\Fg$ in (\ref{24}). Finally, $P^L_{ij}$ and $P^R_A$ are the left- and right-moving momenta of a $\Gamma^{(6,22)}$ Narain-lattice describing the compactification of the heterotic string on the $T^6$ torus. They encode the full dependence of the amplitude on the corresponding $6\times 22=132$ moduli, which form the manifold
\begin{align}
\mathcal{M}=\frac{SO(6,22)}{SO(6)\times SO(22)},
\end{align}
as explained in Section \ref{subcoset}. The exact parameterization of the lattice momenta, however, will be of no importance to our calculations and would involve the explicit construction of the world-sheet sigma model action, starting from the four-dimensional action of $\cN=4$ supergravity coupled to 22 vector multiplets. The left-moving momenta $P^L_{ij}$ are formulated in a complex $SU(4)$ basis and their square is given by
\begin{align}
(P^L)^2=\frac{1}{8}\epsilon^{ijkl}P^L_{ij}P^L_{kl},
\end{align}
which is manifestly real and $SU(4)$ invariant. Moreover, in order to streamline our notation, we will also introduce the following projection of the momenta
\begin{align}
\uPu\equiv \frac{1}{2}\upp^{ij}P^L_{ij}.\label{momentprojection}
\end{align}
\section{First-order harmonicity relation}\label{Sect:FirstOrder}
With the above setting, we are now in a position to discuss the harmonicity equation (\ref{25}) (or (\ref{covfirstorderder})). In \cite{Antoniadis:2006mr}, it was shown that $\Fg$ satisfy the following relation
\begin{align}
\epsilon^{ijkl} \frac{\partial}{\partial \ua ^i} \frac{\partial}{\partial \ub ^j}D_{kl,A}\Fg=0,\label{WeakFirstOrder}
\end{align}
up to an anomaly, which was calculated explicitly. The action of the differential $D_{ij,A}$ with respect to the moduli $\phi_{ij,A}$ can be analyzed in two different ways:
\begin{itemize}
\item From the world-sheet point of view, it amounts inserting the scalar vertex operator
\begin{align}
V^{\text{mod.}}_{ij,A}=-\frac{1}{2\pi}\partial X_{ij}\bar{J}_A(z)e^{ip\cdot X},\label{scalarworldsheetvertex}
\end{align}
into the correlation function, where $X_{ij}$ are the internal bosonic coordinates in an $SU(4)$ basis, satisfying the pseudo-reality condition
\begin{align}
\overline{X^{ij}}=\frac{1}{2}\epsilon^{ijkl}X_{kl},
\end{align}
and $\bar{J}_A$ are the right-moving (Abelian) currents.

This approach is rather cumbersome, since the correlator corresponding to e.g. (\ref{WeakFirstOrder}) contains $(2g+3)$ vertices, for which all possible contractions need to be considered. We will therefore rather resort to the following approach.
\item In terms of the $\Gamma^{(6,22)}$ lattice momenta, the differentials act as infinitesimal Lorentz boosts\footnote{Note a factor of 2 misprint in eq.~(10.7) of \cite{Antoniadis:2006mr} which had no effect in the subsequent analysis.}
\begin{align}
&D_{ij,A}P^L_{kl}=\epsilon_{ijkl}P^R_A, &&D_{ij,A}P^R_B=\frac{\delta_{AB}}{2}P^L_{ij}.\label{naiveRules}
\end{align}
These rules were proved in \cite{Antoniadis:2006mr} by an explicit world-sheet computation at the linearized level. It can be easily checked that they in fact reproduce the algebra (\ref{algebra}), up to normalization factors. Moreover, they annihilate the $SO(6,22)$-square of the lattice vectors
\begin{align}
D_{ij,A}\left((P^L)^2-(P^R)^2\right)=0.
\end{align}
\end{itemize}
As we have seen in Section \ref{global}, the general harmonic expansion of $\Fg$ suggests that (\ref{WeakFirstOrder}) is in fact merely a consequence of the stronger relation (\ref{25}). The goal of this Section is to explicitly test the validity of (\ref{25}) and to examine whether its right hand side is modified by an anomaly as it was the case for (\ref{WeakFirstOrder}).

The computation is done in a straight-forward way using the differentiation rules (\ref{naiveRules})
\begin{align}
E_1\equiv&\epsilon_{ab}\epsilon^{ijkl}\frac{\partial}{\partial \ubb ^j}D_{kl,A}\mathcal{F}_g^{(\text{HET})}=\label{weakstart}\\
&=\epsilon_{ab}\epsilon^{ijkl}\frac{\partial}{\partial \ubb^j}D_{kl,A}\int \frac{d^2\tau}{\bar{\eta}^{24}}\tau_2^{2g-1}G_{g+1}\sum_{(P^L,P^R)}(\uPu )^{2g-2}q^{\frac{1}{2}(P^L)^2}\bar{q}^{\frac{1}{2}(P^R)^2}=\nonumber\\
&=2\epsilon_{ab}\frac{\partial}{\partial \ubb^j}\int \frac{d^2\tau}{\bar{\eta}^{24}}\tau_2^{2g-1}G_{g+1}\sum_{(P^L,P^R)}(\uPu)^{2g-3}\bigg[(2g-2)\upp^{ij}-\pi \tau_2\epsilon^{ijkl}P^L_{kl}(\uPu )\bigg]\cdot\nonumber\\
&\hspace{1cm}\cdot P_A^Rq^{\frac{1}{2}(P^L)^2}\bar{q}^{\frac{1}{2}(P^R)^2},\nonumber
\end{align}
which was essentially already found in \cite{Antoniadis:2006mr}. Using the simple identity
\begin{align}
\epsilon_{ab}\frac{\partial}{\partial \ubb^j}P^L_{++}=\uaa^iP^L_{ij},
\end{align}
we can easily calculate the harmonic partial derivative
\begin{align}
E_1=2\int \frac{d^2\tau}{\bar{\eta}^{24}}&\tau_2^{2g-1}G_{g+1}\sum_{(P^L,P^R)}\bigg[3(2g-2)\uaa^i(\uPu)-(2g-2)\pi \tau_2\uaa^mP^L_{mj}\epsilon^{ijkl}P^L_{kl}(\uPu)+\nonumber\\
&+(2g-3)(2g-2)\uaa^mP^L_{mj}\upp^{ij}\bigg](\uPu )^{2g-4}P_A^Rq^{\frac{1}{2}(P^L)^2}\bar{q}^{\frac{1}{2}(P^R)^2}.\nonumber
\end{align}
Using furthermore the trivial relation
\begin{align}
&\epsilon^{inkl}P^L_{jn}P^{L}_{kl}=2\delta_j^i(P^L)^2,\label{momentidentity}
\end{align}
we can further simplify the expression
\begin{align}
E_1&=2(2g-2)\uaa^i\int \frac{d^2\tau}{\bar{\eta}^{24}}G_{g+1}\sum_{(P^L,P^R)}\left[2g\tau_2^{2g-1}-2\pi \tau_2^{2g}(P^L)^2\right](\uPu )^{2g-3}\cdot\nonumber\\
&\hspace{2cm}\cdot P_A^Rq^{\frac{1}{2}(P^L)^2}\bar{q}^{\frac{1}{2}(P^R)^2}=\nonumber\\
&=4i(2g-2)\uaa^i\int \frac{d^2\tau}{\bar{\eta}^{24}}G_{g+1}\sum_{(P^L,P^R)}\frac{\partial}{\partial\tau}\left[\tau^{2g}(\uPu )^{2g-3}P_A^Rq^{\frac{1}{2}(P^L)^2}\bar{q}^{\frac{1}{2}(P^R)^2}\right].\nonumber
\end{align}
At this point, we perform a partial integration in $\tau$ and use modular invariance together with the exponential suppression in the infra-red region $\tau_2\to\infty$, due to the presence of $P_L$ for $g > 1$, to conclude that there are no boundary terms we have to worry about.\footnote{Note that the equation is trivially fulfilled in the case $g=1$, since $\mathcal{F}_1^{(\text{HET})}$ is independent of the harmonic variables.} The only contribution therefore comes when the $\tau$-derivative acts on $G_{g+1}$. Using the identity (\ref{Ggeq}) we get
\begin{align}
E_1=-2(2g-2)\pi \uaa ^i\int \frac{d^2\tau}{\bar{\eta}^{24}}\tau_2^{2g-2}G_{g}\sum_{(P^L,P^R)}(\uPu )^{2g-3}P_A^Rq^{\frac{1}{2}(P^L)^2}\bar{q}^{\frac{1}{2}(P^R)^2}.\label{resultweak2}
\end{align}

This result has to be contrasted with the expression
\begin{align}
&D_{++,A}\mathcal{F}_{g-1}^{(\text{HET})},
\end{align}
where we have used the same projection as in (\ref{momentprojection})
\begin{align}
D_{++,A}\equiv \frac{1}{2}\bar{u}_{++}^{ij}D_{ij,A}.
\end{align}
The calculation follows much along the same lines as before and yields the result
\begin{align}
&D_{++,A}\mathcal{F}_{g-1}^{(\text{HET})}=-2\pi\int \frac{d^2\tau}{\bar{\eta}^{24}}\tau^{2g-2}_2G_{g}\sum_{(P^L,P^R)}(\uPu )^{2g-3}P_A^Rq^{\frac{1}{2}(P^L)^2}\bar{q}^{\frac{1}{2}(P^R)^2}.
\end{align}
Comparing this result with (\ref{resultweak2}), one concludes
\begin{align}
\epsilon_{ab}\epsilon^{ijkl}\frac{\partial}{\partial \ubb ^j}D_{kl,A}\Fg=(2g-2)\uaa^iD_{++,A}\mathcal{F}_{g-1}^{(\text{HET})}.\label{weakharmcov}
\end{align}
Since the $\mathcal{F}_{g-1}^{(\text{HET})}$, which appears on the right hand side is of lower order in $g$ than the initial one we considered on the left hand side, this term can be interpreted as an anomaly to the harmonicity relation. This is justified by comparison to the holomorphic anomaly equation \cite{Antoniadis:1993ze}-\cite{Bershadsky:1993ta}, where (for the type II theory) the lower genus\footnote{In \cite{Bershadsky:1993cx} scattering amplitudes of two (self-dual) Riemann tensors and $(2g-2)$ graviphoton field strengths in type~II theory compactified on Calabi-Yau threefolds were considered. In these amplitudes, the number $g$ corresponds to the genus of the world-sheet Riemann surface.} terms have their origin from boundary contributions in the moduli space of genus $g$ world-sheets.

As a trivial consistency check of this result, we can try to recover the weaker harmonicity relation presented in \cite{Antoniadis:2006mr}, by applying a second partial differentiation with respect to $\uaa ^i$ to (\ref{weakharmcov}) using the fact that it commutes with $\frac{\partial}{\partial \ubb ^j}D_{kl,A}$
\begin{align}
&\epsilon_{ab}\epsilon^{ijkl}\frac{\partial}{\partial \uaa ^i}\frac{\partial}{\partial \ubb ^j}D_{kl,A}\Fg=2(2g-2)(2g+1)D_{++,A}\mathcal{F}_{g-1}^{(\text{HET})},\label{strongharm}
\end{align}
which is precisely the result found in \cite{Antoniadis:2006mr}.
\section{Second-order constraint}\label{Sect:SecondOrderRelation}
In the same way as equation (\ref{25}), we can now check relation (\ref{25'}) (or rather its counterparts (\ref{conts}) and (\ref{commu}) taking into account the curvature of the moduli space) by directly applying the corresponding differential operator to the topological amplitude $\Fg$. We use again the differentiation rules (\ref{naiveRules}) to obtain
{\allowdisplaybreaks
\begin{align}
E_2&\equiv\epsilon^{ijkm}D_{ij,A}D_{kl,B}\Fg=\nonumber\\
&=\epsilon^{ijkm}D_{ij,A}\int\frac{d^2\tau}{\bar{\eta}^{24}}\tau_2^{2g-1}G_{g+1}\sum_{(P^L,P^R)}\bigg[\frac{1}{2}(2g-2)\epsilon_{klpq}\upp^{pq}-2\pi\tau_2P^L_{kl}(\uPu)\bigg](\uPu)^{2g-3}\cdot\nonumber\\
&\hspace{1cm}\cdot P_B^Rq^{\frac{1}{2}(P^L)^2}\bar{q}^{\frac{1}{2}(P^R)^2}.\nonumber
\end{align}
Taking now the second scalar derivative, one has
\begin{align}
E_2&=\epsilon^{ijkm}\int\frac{d^2\tau}{\bar{\eta}^{24}}\tau_2^{2g-1}G_{g+1}\sum_{(P^L,P^R)}\bigg\{\left[-2\pi\tau_2\epsilon_{ijkl}(\uPu)-\pi\tau_2P^L_{kl}\epsilon_{ijpq}\upp^{pq}\right](\uPu)^{2g-3}+\nonumber\\
&\hspace{0.7cm}+\left[\frac{1}{2}(2g-2)\epsilon_{klpq}\upp^{pq}-2\pi\tau_2P^L_{kl}(\uPu)\right]\cdot\left[\frac{1}{2}(2g-3)\epsilon_{ijrs}\upp^{rs}-2\pi\tau_2P^L_{ij}(\uPu)\right]\cdot\nonumber\\
&\hspace{0.7cm}\cdot(\uPu)^{2g-4}\bigg\}\cdot P^R_AP^R_Bq^{\frac{1}{2}(P^L)^2}\bar{q}^{\frac{1}{2}(P^R)^2}+\nonumber\\
&\hspace{0.5cm}+\frac{\delta_{AB}}{2}\epsilon^{ijkm}\int\frac{d^2\tau}{\bar{\eta}^{24}}\tau_2^{2g-1}G_{g+1}\sum_{(P^L,P^R)}P^L_{ij}\left[\frac{1}{2}(2g-2)\epsilon_{klpq}\upp^{pq}-2\pi \tau_2P^L_{kl}(\uPu)\right]\cdot\nonumber\\
&\hspace{0.7cm}\cdot(\uPu)^{2g-3}q^{\frac{1}{2}(P^L)^2}\bar{q}^{\frac{1}{2}(P^R)^2}.\nonumber
\end{align}}
Regrouping the terms furthermore
\begin{align}
E_2&=\int\frac{d^2\tau}{\bar{\eta}^{24}}\tau_2^{2g-1}G_{g+1}\sum_{(P^L,P^R)}\bigg\{-12\pi\tau_2\delta_l^m(\uPu)^{2g-2}-4\pi(2g-2)\tau_2P^L_{kl}\upp^{km}(\uPu)^{2g-3}-\nonumber\\
&\hspace{1cm}-\pi\tau_2(2g-2)\epsilon_{klpq}\epsilon^{ijkm}\upp^{pq}P^L_{ij}(\uPu)^{2g-3}+4\pi^2\tau_2^2\epsilon^{ijkm}P^L_{kl}P^L_{ij}(\uPu)^{2g-2}\bigg\}\cdot\nonumber\\
&\hspace{1cm}\cdot P^R_AP^R_Bq^{\frac{1}{2}(P^L)^2}\bar{q}^{\frac{1}{2}(P^R)^2}+\nonumber\\
&\hspace{0.5cm}+\frac{\delta_{AB}}{2}\left[(2g-2)\epsilon^{ijkm}\epsilon_{klpq}P^L_{ij}\upp^{pq}-2\pi\tau_2\epsilon^{ijkm}P^L_{ij}P^L_{kl}(\uPu)\right](\uPu)^{2g-3}q^{\frac{1}{2}(P^L)^2}\bar{q}^{\frac{1}{2}(P^R)^2}\nonumber
\end{align}
and using the relations
\begin{align}
&\epsilon^{klmi}P^L_{kl}P^L_{mj}=2\delta_j^i(P^L)^2,\\
&\frac{1}{2}\epsilon^{klmi}\epsilon_{mjpq}P^L_{kl}\upp^{pq}+2P^L_{pj}\upp^{pi}=2(\uPu )\delta_j^i,
\end{align}
we obtain
\begin{align}
&E_2=\!-4\pi \int\frac{d^2\tau}{\bar{\eta}^{24}}\tau_2^{2g}G_{g+1}\sum_{(P^L,P^R)}\left[(2g+1)-2\pi\tau_2(P^L)^2\right]\delta_l^m(\uPu)^{2g-2}P^R_AP^R_Bq^{\frac{1}{2}(P^L)^2}\bar{q}^{\frac{1}{2}(P^R)^2}+\nonumber\\
&\hspace{0.5cm}+\delta_{AB}\int\frac{d^2\tau}{\bar{\eta}^{24}}\tau_2^{2g-1}G_{g+1}\sum_{(P^L,P^R)}\left[2g-2\pi\tau_2(P^L)^2\right]\delta_m^l(\uPu)^{2g-2}q^{\frac{1}{2}(P^L)^2}\bar{q}^{\frac{1}{2}(P^R)^2}-\nonumber\\
&\hspace{0.5cm}-\delta_{AB}\int\frac{d^2\tau}{\bar{\eta}^{24}}\tau_2^{2g-1}G_{g+1}\sum_{(P^L,P^R)}\left[2(\uPu)\delta_m^l+(2g-2)P^L_{pm}u_{++}^{pl}\right](\uPu)^{2g-3}q^{\frac{1}{2}(P^L)^2}\bar{q}^{\frac{1}{2}(P^R)^2}
\end{align}
At this point one can check that the first two lines as well as the last line, separately, are indeed modular invariant (for the last line, this follows mainly from the presence of the harmonics). Moreover, the first two lines can be written as differentiations with respect to the torus Teichm\"uller parameter:
\begin{align}
&E_2=-8i\pi \delta_l^m\int\frac{d^2\tau}{\bar{\eta}^{24}}G_{g+1}\sum_{(P^L,P^R)}\frac{\partial}{\partial\tau}\left[\tau_2^{2g+1}(\uPu)^{2g-2}P^R_AP^R_Bq^{\frac{1}{2}(P^L)^2}\bar{q}^{\frac{1}{2}(P^R)^2}\right]+\nonumber\\
&\hspace{0.5cm}+2i\delta_{AB}\delta_l^m\int\frac{d^2\tau}{\bar{\eta}^{24}}G_{g+1}\sum_{(P^L,P^R)}\frac{\partial}{\partial\tau}\left[\tau_2^{2g}(\uPu)^{2g-2}q^{\frac{1}{2}(P^L)^2}\bar{q}^{\frac{1}{2}(P^R)^2}\right]-\nonumber\\
&\hspace{0.5cm}-\delta_{AB}\int\frac{d^2\tau}{\bar{\eta}^{24}}\tau_2^{2g-1}G_{g+1}\sum_{(P^L,P^R)}\left[2(\uPu)\delta_m^l+(2g-2)P^L_{pm}u_{++}^{pl}\right](\uPu)^{2g-3}q^{\frac{1}{2}(P^L)^2}\bar{q}^{\frac{1}{2}(P^R)^2}.
\end{align}
Since these terms are modular invariant, one is allowed to perform a partial integration, with $\frac{\partial}{\partial\tau}$ only hitting the factor $G_{g+1}$
\begin{align}
&E_2=4\pi^2\delta_l^m\int\frac{d^2\tau}{\bar{\eta}^{24}}\tau_2^{2g-1}G_g\sum_{(P^L,P^R)}(\uPu)^{2g-2}P^R_AP^R_Bq^{\frac{1}{2}(P^L)^2}\bar{q}^{\frac{1}{2}(P^R)^2}-\nonumber\\
&\hspace{0.5cm}-\delta_{AB}\pi\delta_l^m\int\frac{d^2\tau}{\bar{\eta}^{24}}\tau_2^{2g}G_{g}\sum_{(P^L,P^R)}(\uPu)^{2g-2}q^{\frac{1}{2}(P^L)^2}\bar{q}^{\frac{1}{2}(P^R)^2}-\nonumber\\
&\hspace{0.5cm}-\delta_{AB}\int\frac{d^2\tau}{\bar{\eta}^{24}}\tau_2^{2g-1}G_{g+1}\sum_{(P^L,P^R)}\left[2(\uPu)\delta_m^l+(2g-2)P^L_{pm}u_{++}^{pl}\right](\uPu)^{2g-3}q^{\frac{1}{2}(P^L)^2}\bar{q}^{\frac{1}{2}(P^R)^2}.\nonumber
\end{align}

This expression can be contrasted with
\begin{align}
D_{++,A}D_{++,B}\mathcal{F}_{g-1}^{(\text{HET})},
\end{align}
which can be computed using exactly the same rules as before
\begin{align}
&D_{++,A}D_{++,B}\mathcal{F}_{g-1}^{(\text{HET})}=\nonumber\\
&=-2\pi D_{++,A}\int \frac{d^2\tau}{\bar{\eta}^{24}}\tau_2^{2g-2}G_{g}\sum_{(P^L,P^R)}(\uPu)^{2g-3}P^R_Bq^{\frac{1}{2}(P^L)^2}\bar{q}^{\frac{1}{2}(P^R)^2}=\nonumber\\
&=4\pi^2\int \frac{d^2\tau}{\bar{\eta}^{24}}\tau_2^{2g-1}G_{g}\sum_{(P^L,P^R)}(\uPu)^{2g-2}P^R_AP^R_Bq^{\frac{1}{2}(P^L)^2}\bar{q}^{\frac{1}{2}(P^R)^2}-\nonumber\\*
&\hspace{0.5cm}-\pi \delta_{AB}\int \frac{d^2\tau}{\bar{\eta}^{24}}\tau_2^{2g-2}G_{g}\sum_{(P^L,P^R)}(\uPu)^{2g-2}q^{\frac{1}{2}(P^L)^2}\bar{q}^{\frac{1}{2}(P^R)^2}.\nonumber
\end{align}
From this, we conclude for the second order constraint
\begin{align}
&\epsilon^{ijkm}D_{ij,A}D_{kl,B}\Fg=\delta^m_lD_{++,A}D_{++,B}\mathcal{F}_{g-1}^{(\text{HET})}-2\delta^m_l\delta_{AB}\Fg-\nonumber\\
&\hspace{0.5cm}-(2g-2)\delta_{AB}\int\frac{d^2\tau}{\bar{\eta}^{24}}\tau_2^{2g-1}G_{g+1}\sum_{(P^L,P^R)}P^L_{pm}\upp^{pl}(\uPu)^{2g-3}q^{\frac{1}{2}(P^L)^2}\bar{q}^{\frac{1}{2}(P^R)^2}.\label{striresrot}
\end{align}
Notice that the last two terms are not of the form of an anomaly but are generic ``hard" contributions to the equation. As one can easily see, they correspond however to an $SU(4)$($\sim SO(6)$) rotation acting on the harmonics inside $\mathcal{F}_{g}^{(\text{HET})}$, which is exactly what one expects according to (\ref{commu}).

On the other hand, (\ref{striresrot}) is vastly simplified when we contract its free $SU(4)$ indices:
\begin{align}
\epsilon^{ijkl}D_{ij,A}D_{kl,B}\Fg=4D_{++,A}D_{++,B}\mathcal{F}_{g-1}^{(\text{HET})}-4(g+1)\delta_{AB}\Fg.\label{secondorderstring}
\end{align}
Comparing this result to (\ref{conts}), we conclude that besides the anomalous term proportional to $\mathcal{F}_{g-1}^{(\text{HET})}$ the two relations indeed agree, up to an irrelevant normalization.\\

Finally, let us mention in passing that a second order differentiation, which is antisymmetrized in the $SO(22)$ indices, is exactly vanishing
\begin{align}
\frac{\partial}{\partial\phi_{[A}^{ij}}\frac{\partial}{\partial\phi_{B]}^{kl}}\Fg=0.\label{trivialantisymm}
\end{align}
This can be seen most easily by representing the above expression as a correlator with two additional scalar vertices inserted and realizing that its right-moving part is given by
\begin{align}
\langle\bar{J}_{[A}(\bar{z})\bar{J}_{B]}(\bar{w})\rangle,\label{rightmovcorr}
\end{align}
which follows from the form of the scalar vertex operator (\ref{scalarworldsheetvertex}). Since the right-moving currents are Abelian, it follows that expression (\ref{rightmovcorr}) is identically zero. Note in particular that in this case there is not even an anomaly, and (\ref{trivialantisymm}) remains in fact exact at the quantum level.
\section{Harmonicity in six dimensions}\label{sixdim}

\subsection{The origin of the harmonicity constraint}\label{Sect:Origin6dimensions}

In this subsection we summarize a few key points about six-dimensional harmonic superspace and derive the corresponding  harmonicity constraint. The discussion closely follows that of the $\cN=4$ case in four dimensions, therefore it is very brief.

We consider ${\cal N}=(1,1)$ supersymmetry in six dimensions whose automorphism group is $SU(2)_L\times SU(2)_R$. Let us introduce harmonic variables $v^I_a$ for $SU(2)_L$ and  $v^{{\dot I}}_{{\dot a}}$ for $SU(2)_R$, together with their conjugates $v^a_I = (v^I_a)^*$ and  $v^{{\dot a}}_{{\dot I}} = (v^{{\dot I}}_{{\dot a}})^*$. Here $a, {\dot a}$ are $SU(2)$ doublet indices while $I,{\dot I}=1,2$ are projections onto the subgroup  $U(1)_L\times U(1)_R$. They satisfy the completeness conditions
\begin{equation}\label{1}
    v^I_a\, v^a_J = \delta^I_J\,, \quad v^a_I\, v^I_b = \delta^a_b
\end{equation}
(and similarly for $v^{{\dot I}}_{{\dot a}}$). Raising and lowering the indices with $\ep_{ab}$, $\ep_{IJ}$, etc., we can rewrite the non-trivial part of (\ref{1}) as the unit determinant condition
\begin{equation}\label{1'}
    \ep^{ab} v^1_a v^2_b = 1\ .
\end{equation}
In fact, the harmonics can be viewed as matrices of the corresponding $SU(2)$ groups.

The harmonic functions are supposed to have harmonic expansions homogeneous under the action of the subgroup $U(1)_L\times U(1)_R$. For example, a function of unit  $U(1)_L\times U(1)_R$ charges has the expansion
\begin{equation}\label{2}
    \phi^{1{\dot 1}}(v) = \phi^{a{\dot a}} v^1_a v^{{\dot 1}}_{{\dot a}} + \phi^{(abc){\dot a}} v^1_a v^1_b v^2_c v^{{\dot 1}}_{{\dot a}} + \phi^{a({\dot a}{\dot b}\dot c)} v^1_a v^{{\dot 1}}_{{\dot a}} v^{{\dot 1}}_{{\dot b}} v^{{\dot 2}}_{\dot c} + \cdots\ ,
\end{equation}
so that in each term the number of $v^1$ exceeds by one the number of $v^2$ (the same for $v^{\dot 1, \dot 2}$). Notice that due to the constraint (\ref{1'}) each component is an irrep of $SU(2)_L\times SU(2)_R$ (i.e., only symmetrized indices appear). Effectively, such homogeneous functions live on the coset $S^2_L\times S^2_R = (SU(2)_L/U(1)_L)\times (SU(2)_R/U(1)_R)$.

The introduction of harmonic variables allows us to define G-analytic superfields which depend only on half of the Grassmann variables,\footnote{We use $SU^*(4)\sim SO(1,5)$ chiral spinor notation with left-handed $\psi_\a$ and right-handed $\bar\psi^\a$ spinors.} e.g. on $\q^1_{\a} = v^1_a \q^a_{ \a}$, $ \bq^\a_{ \dot 2} =  v^{\ad}_{\dot 2} \bq^\a_{\ad} = \bq^{\a \dot 1}$. One such short superfield describes the (on-shell) vector multiplet
\begin{equation}
  Y^{1{\dot 1}}(\q^1, \bq_{\dot 2}, v) = \phi^{a{\dot a}} v^1_a v^{{\dot 1}}_{{\dot a}} + \q^1_\a  \bar\psi^{{\a}}_{\ad} v_{{\dot 2}}^{{\dot a}}
  + \bq_{{\dot 2}}^\a \psi_\a^a v^1_a + \bq_{{\dot 2}}\sigma^{\m\n}\q^1 \, F_{\m\n}  + \cdots  \ . \label{3}
\end{equation}
Notice the conservation of the overall charges $1,{\dot 1}$  carried by the projected Grassmann variables or by the explicit harmonics projecting the component fields. This superfield is real in the sense $\widetilde{Y^{1{\dot 1}}} = Y^{1{\dot 1}}$, where $\widetilde{}\ $ is a combination of complex conjugation with a reflection on $S^2\times S^2$ preserving G-analyticity. In particular, this implies the reality of the first component, $(\phi^{a{\dot a}} )^* = \ep_{ab} \ep_{{\dot a}{\dot b}} \phi^{b{\dot b}} $.

Another short superfield of the same type describes the (on-shell) Weyl multiplet \cite{Berkovits:1994vy}
\begin{equation}\label{4}
    (W^{1{\dot 1}})_\a{}^\b(\q^1, \bq_{\dot 2}, v) = (T^{a{\dot a}})_\a{}^\b v^1_a v^{{\dot 1}}_{{\dot a}}  + \q^1_\gamma \bq^\delta_{{\dot 2}} R_{\a\delta}{}^{\b\gamma} + \cdots \ ,
\end{equation}
where $(W^{1{\dot 1}})_\a{}^\b $ is in the adjoint of $SU^*(4)$ ($(W^{1{\dot 1}})_\a{}^\a = 0$), $(T^{a{\dot a}})_\a{}^\b = (T^{a{\dot a}})_{\m\n} (\sigma^{\m\n})_\a{}^\b$ are the graviphoton field strengths and $R_{\a\delta}{}^{\b\gamma} = R_{\m\n\lambda\rho} (\sigma^{\m\n})_\a{}^\b (\sigma^{\lambda\rho})_\delta{}^\gamma$ is the curvature.

In \cite{Berkovits:1994vy} the following term of the six-dimensional effective action was considered:
\begin{align}
    \int d^6x \ dv \ d^4\q^1 \ d^4\bq_{\dot 2}\ &\left[\ep^{\a_1\a_2\a_3\a_4} \ep_{\b_1\b_2\b_3\b_4}(W^{1{\dot 1}})_{\a_1}{}^{\b_1} (W^{1{\dot 1}})_{\a_2}{}^{\b_2}(W^{1{\dot 1}})_{\a_3}{}^{\b_3}(W^{1{\dot 1}})_{\a_4}{}^{\b_4} \right]^g \nn\\
     & \times F_{(1)^{4g-4}  {({\dot 1})}^{4g-4}} ( Y^{1{\dot 1}}, v)\ . \label{5}
\end{align}
In fact, what appears in (\ref{5}) is the determinant of the $4\times4$ traceless matrix $(W^{1{\dot 1}})_\a{}^\b$. This is a Lorentz invariant which breaks up into two independent invariants, $[{\rm Tr} (W^{1{\dot 1}})^2]^2$ and ${\rm Tr} (W^{1{\dot 1}})^4$. We could use anyone of them to construct an effective action term similar to (\ref{5}). However, upon decompactification of the four-dimensional couplings (\ref{19}) and (\ref{19'}), one can show that only the first of the two invariants contributes. We will eventually study this case in the next subsection. The corresponding effective action term is
\begin{equation}\label{effact6}
    \int d^6x \ dv \ d^4\q^1 \ d^4\bq_{\dot 2}\ \left[(W^{1{\dot 1}})_{\a}{}^{\b} (W^{1{\dot 1}})_{\b}{}^{\a} \right]^{g+1}\ F_{(1)^{2g-2}  {({\dot 1})}^{2g-2}} ( Y^{1{\dot 1}}, v)\ .
\end{equation}

The function $F_{(1)^{m}  {({\dot 1})}^{m}}$  ($m=2g-2$) has to carry a `negative' (i.e. indices $1,{\dot 1}$ downstairs) charges of each kind, in order to compensate that of the $K$ factor ($+4g$) and of the Grassmann measure $(-4)$. We consider functions of the type
\begin{equation}\label{6}
    F_{(1)^{m}  {({\dot 1})}^{m}} = \sum_{n=0}^\infty  \ \xi_{(1)^{m+n}  {({\dot 1})}^{m+n}}\ (Y^{1{\dot 1}})^n\ ,
\end{equation}
where
\begin{equation}\label{6'}
     \xi_{(1)^{p}  {({\dot 1})}^{p}} = \xi_{(a_1\cdots a_p) ({\dot a}_1\cdots {\dot a}_p)} v_1^{a_1} \cdots v_1^{a_{p}}\  v_{{\dot 1}}^{{\dot a}_1} \cdots v_{{\dot 1}}^{{\dot a}_{p}}
\end{equation}
introduces a set of {constant}  $SU(2)_L\times SU(2)_R$ multispinors, thus explicitly breaking the symmetry.

Let us examine the coupling (\ref{effact6}) in some detail. First of all,  from the term $(W^{1{\dot 1}}W^{1{\dot 1}})^{g+1}$ we only consider contributions of the type
\begin{equation}\label{7}
    (\q^1)^4(\bq_{\dot 2})^4 \ R^4\ (T^{1{\dot 1}})^{m} \ .
\end{equation}
The Grassmann factor saturates the $\q$ integrals. The harmonic dependence comes from the factor
\begin{equation}\label{8}
    (T^{1{\dot 1}})^{m} = T^{(a_1({\dot a}_1} \cdots T^{a_{m}) {\dot a}_{m})} v^1_{a_1} \cdots v^1_{a_{m}}\  v^{{\dot 1}}_{{\dot a}_1} \cdots v^{{\dot 1}}_{{\dot a}_{m}}\ .
\end{equation}
Notice that the projection with commuting harmonic variables forces symmetrization of the indices of the $T$'s. Thus, this term contributes an irrep of each $SU(2)$ of weight $m$. Since the harmonic integral in (\ref{effact6}) only sees the singlet part of the integrand, we have to find a matching irrep in the $F$ sector, so that together they can form a singlet. Let us look at a term from (\ref{6}) (where we replace the superfield $Y$ by its first component $\phi$),
\begin{eqnarray}
  \xi_{(1)^{m+n}  {({\dot 1})}^{m+n}}\ (\phi^{1{\dot 1}})^n &=& \xi_{(a_1\cdots a_{m+n}) ({\dot a}_1\cdots {\dot a}_{m+n})} \ v_1^{a_1} \cdots v_1^{a_{m+n}}\  v_{{\dot 1}}^{{\dot a}_1} \cdots v_{{\dot 1}}^{{\dot a}_{m+n}} \nonumber\\
  && \times  \phi^{(b_1({\dot b}_1} \cdots \phi^{b_{n}) {\dot b}_{n})} v^1_{b_1} \cdots v^1_{b_{n}}\  v^{{\dot 1}}_{{\dot b}_1} \cdots v^{{\dot 1}}_{{\dot b}_{n}}\ . \label{9}
\end{eqnarray}
The first factor involves only harmonics with upper $SU(2)$ indices, the second only with lower indices. Such products of harmonics are reducible. Using the defining conditions (\ref{1}), we can  decompose a reducible product of  $v^1_a$ with $v^b_1$ as follows: $v^1_a v^b_1 = 1/2\, \delta^b_a + v^1_{\{a} v^{b\}}_1$,  where $\{\}$ denotes the traceless part. Contracting the indices of all $v^1$'s and $v^{{\dot 1}}$'s with those of a subset of the $v_1$'s and $v_{\dot{1}}$'s, we can eliminate the $v^1$'s and $v^{{\dot 1}}$'s from (\ref{9}). The result is the irrep of weight $m$ of each $SU(2)$ needed to match that in (\ref{8}); any traceless combination $v^1_{\{i} v^{j\}}_1$ will contribute to an irrep of higher isospin without a match in  (\ref{8}), thus irrelevant for the harmonic integral. So, we can reduce (\ref{6}) to its relevant part
\begin{equation}\label{10}
    {\cal F} = \sum_n  \ \xi_{(a_1\cdots a_{m+n}) ({\dot a}_1\cdots {\dot a}_{m+n})}  \ v_1^{a_1} \cdots v_1^{a_{m}}\  v_{{\dot 1}}^{{\dot a}_1} \cdots v_{{\dot 1}}^{{\dot a}_{m}} \ \phi^{a_{m+1}{\dot a}_{m+1}} \cdots \phi^{a_{m+n} {\dot a}_{m+n}}\ .
\end{equation}

It is important to realize that the $\xi$ tensor in (\ref{10}) has all its indices symmetrized. This is the origin of the harmonicity constraint
\begin{equation}\label{11}
    \ep^{ab}\frac{\pa}{\pa v^a_1} \frac{\pa}{\pa \phi^{b\dot b}}\ {\cal F} = 0\ .
\end{equation}
It involves a partial derivative with respect to $v_1$. Strictly speaking, such an operation is illegal in the harmonic formalism, since the variables $v_1$ and $v^1$ are not independent, as can be seen from (\ref{1}). However, in  (\ref{10}) there are only $v_1$'s left, so we can {formally} take such a derivative. In fact, if needed, (\ref{11}) can also be expressed using covariant harmonic derivatives as in (\ref{25prime}).

In principle, we could go on and discuss the coset space $SO(4,n)/SO(4)\times SO(n)$ parametrized by the scalars $\phi$ of the vector multiplets (\ref{3}) in a manner similar to that of Sect. \ref{coset}. The conclusion would be a second-order constraint analogous to (\ref{correctcons}). However, in six dimensions we do not have the setup of conformal supergravity of Sect. \ref{css} which allowed us to fix the value of the charge $Z_0$ in (\ref{conts}). Therefore, we can make a prediction for the structure of this constraint, but we cannot explain the precise value of the coefficient obtained from the string calculation, see (\ref{strire}).

\subsection{Decompactification of four-dimensional amplitudes }
\subsubsection{Decompactification limit}

In order to round up the six-dimensional discussion, let us now check the field theory predictions by direct string calculations for the decompactification of the topological amplitude (\ref{hetamp}) from four to six dimensions, which corresponds to the coupling (\ref{effact6}). Essentially, it was already shown in \cite{Antoniadis:2006mr} that upon decomposing $T^6$ into $T^4\times T^2$ and the subsequent reduction of the $\Gamma^{(6,22)}$ lattice into
\begin{align}
\Gamma^{(6,22)}\to\Gamma^{(4,20)}\times \Gamma^{(2,2)},\label{latticeReduction}
\end{align}
the weaker version of the first order harmonicity relation (\ref{strongharm}) is reduced to a relation for type II string theory compactified on $K3$, proved in \cite{Ooguri:1995cp}. Below, we will check the stronger relation (\ref{11}) and compute its corresponding quantum anomaly.

In order to perform the reduction (\ref{latticeReduction}) we choose as in \cite{Antoniadis:2006mr} $P^L_{13}$ and its complex conjugate $P^L_{24}$ of $\Gamma^{(6,22)}$ to be entirely in $\Gamma^{(2,2)}$ and the remaining four $P^L_{12}$, $P^L_{14}$ and their complex conjugates $P^L_{34}$, $P^L_{23}$ to form the $\Gamma^{(4,20)}$. In this way, the group $SU(4)$ is reduced to its subgroup $SU(2)_L\times SU(2)_R$ where $SU(2)_L$ and $SU(2)_R$ are acting on the indices $(1,3)$ and $(2,4)$, respectively. In the decompactification limit, $P_{13}$ and $P_{24}$ decouple and are dropped from the correlation function. In this way, $\Gamma^{(4,20)}$ lattice vectors are denoted by:
\begin{align}
P^{(4,20),L}_{a\dot{b}}\ ,\ P^{(4,20),R}_A\hspace{2cm}\text{with}\ a,\dot{b}=1,2
\end{align}
and the index of the right-moving lattice momenta takes now the values $A=1,\ldots,20$. Moreover, the square of the left-moving momenta will be denoted by
\begin{align}
(P^{(4,20),L})^2= \frac{1}{2}P^{(4,20),L}_{a_1\dot{b}_1}P^{(4,20),L}_{a_2\dot{b}_2}\epsilon^{a_1a_2}\epsilon^{\dot{b}_1\dot{b}_2}\, .
\end{align}
In order to make contact with the six-dimensional harmonic coordinates introduced in Section \ref{Sect:Origin6dimensions} we can assemble part of the $SU(4)$ harmonics $\ua$ and $\ub$ into the harmonics of $SU(2)_L\times SU(2)_R$ with the identification
\begin{align}
&(\ua^i, \bar{u}_{-1}^i)\rightarrow (v_1^{a}, v_2^{a}),\\
&(\ub^i, \bar{u}_{-2}^i) \rightarrow (v_{\dot{1}}^{\dot{a}}, v_{\dot{2}}^{\dot{a}}),
\end{align}
and we recall that the $SU(2)_L\times SU(2)_R$ harmonics satisfy the completeness condition (\ref{1}) and the unit determinant condition (\ref{1'}).
Finally, the 1-loop heterotic amplitude (\ref{hetamp}) was shown in \cite{Antoniadis:2006mr} to take the following form after the decompactification of the $T^2$
\begin{align}
\mathcal{F}^{\text{dec}}_g\sim\int \frac{d^2\tau}{\bar{\eta}^{24}}\tau_2^{2g-2} G_{g+1}(\tau,\bar{\tau})\sum_{(P^L,P^R)\in \Gamma^{(4,20)}}\left(v_1^aP^{L}_{a\dot{b}}v_{\dot{1}}^{\dot{b}}\right)^{2g-2}q^{\frac{1}{2}(P^L)^2}\bar{q}^{\frac{1}{2}(P^R)^2},
\label{fgdec}
\end{align}
where from now on, we will drop the $(4,20)$ superscript of the lattice momenta and for further convenience, we define the following shorthand notation
\begin{align}
\vp\equiv \left(v_1^aP^{L}_{a\dot{b}}v_{\dot{1}}^{\dot{b}}\right),
\end{align}
similar to the four-dimensional definition (\ref{momentprojection}).

\subsubsection{Harmonicity relation}
We now study the six-dimensional harmonicity relation (\ref{11}):
\begin{align}
\epsilon^{\dot{a}\dot{b}}\frac{\partial}{\partial v_{\dot{1}}^{\dot{a}}}D_{a\bd,A}\mathcal{F}_g^{\text{dec}},\label{left6}
\end{align}
where the covariant derivative $D_{a\dot{b},A}$ is with respect to the moduli forming the $\Gamma^{(4,20)}$ lattice. We can again apply simple rules for the differentials acting on the lattice momenta, similar to (\ref{naiveRules}):
\begin{align}
&D_{a\ad,A}P^L_{b\bd}=\epsilon_{ab}\epsilon_{\ad\bd}P^R_A, &&D_{a\ad,A}P^R_B=\frac{\delta_{AB}}{2}P^L_{a\ad}.\label{naiveRules6D}
\end{align}
The computation can then be performed in the same straight-forward manner as in the four-dimensional case
\begin{align}
E_1^{\text{dec}}\equiv&\epsilon^{\dot{a}\dot{b}}\frac{\partial}{\partial v_{\dot{1}}^{\dot{a}}}D_{a\bd,A}\int \frac{d^2\tau}{\bar{\eta}^{24}}\tau_2^{2g-2}G_{g+1}\sum_{(P^L,P^R)}(\vp)^{2g-2}q^{\frac{1}{2}(P^L)^2}\bar{q}^{\frac{1}{2}(P^R)^2}=\nonumber\\
&=\epsilon^{\dot{a}\dot{b}}\frac{\partial}{\partial v_{\dot{1}}^{\dot{a}}}\int \frac{d^2\tau}{\bar{\eta}^{24}}\tau_2^{2g-2}G_{g+1}\sum_{(P^L,P^R)}\bigg[(2g-2)v_1^cv_{\dot{1}}^{\dot{c}}\epsilon_{ac}\epsilon_{\bd\dc}+\pi i(\tau-\bar{\tau})\vp P^L_{a\dot{b}}\bigg]\cdot\nonumber\\
&\hspace{1cm}\cdot(\vp)^{2g-3}P_A^Rq^{\frac{1}{2}(P^L)^2}\bar{q}^{\frac{1}{2}(P^R)^2}.\nonumber
\end{align}
The derivative with respect to the harmonic variable yields
\begin{align}
&E_1^{\text{dec}}=\epsilon^{\dot{a}\dot{b}}\int \frac{d^2\tau}{\bar{\eta}^{24}}\tau_2^{2g-2}G_{g+1}\sum_{(P^L,P^R)}\bigg\{\bigg[(2g-2)v_1^c\epsilon_{ac}\epsilon_{\bd\ad}-2\pi \tau_2v_1^cP^L_{a\bd}P^L_{c\ad}\bigg](\vp)^{2g-3}+\nonumber\\
&+(2g-3)\bigg[(2g-2)v_1^cv_{\dot{1}}^{\dc}\epsilon_{ac}\epsilon_{\bd\dc}-2\pi\tau_2P^L_{a\bd}(\vp)\bigg]v_1^cP^L_{c\ad}(\vp)^{2g-4}\bigg\}P_A^Rq^{\frac{1}{2}(P^L)^2}\bar{q}^{\frac{1}{2}(P^R)^2}=\nonumber\\
&=-(2g-2)\int \frac{d^2\tau}{\bar{\eta}^{24}}\tau_2^{2g-2}G_{g+1}\sum_{(P^L,P^R)}(\vp)^{2g-3}\bigg[(2g-1)\epsilon_{ac}v_1^c+2\pi\tau_2v_1^cP^L_{c\ad}\epsilon^{\ad\bd}P^L_{a\bd}\bigg]\cdot\nonumber\\
&\hspace{0.5cm}\cdot P_A^Rq^{\frac{1}{2}(P^L)^2}\bar{q}^{\frac{1}{2}(P^R)^2}.\nonumber
\end{align}
We can now make use of the identity
\begin{align}
P^L_{b\dot{a}}\epsilon^{\dot{a}\dot{b}}P^L_{a\dot{b}}=-(P^L)^2\epsilon_{ab},
\end{align}
which simplifies the expression to
\begin{align}
&E_1^{\text{dec}}=\nonumber\\
&\!-(2g-2)\!\int\! \frac{d^2\tau}{\bar{\eta}^{24}}&\tau_2^{2g-2}G_{g+1}\!\!\sum_{(P^L,P^R)}\epsilon_{ac}v_1^c(\vp)^{2g-3}\bigg[(2g-1)-2\pi\tau_2(P^L)^2\bigg] P_A^Rq^{\frac{1}{2}(P^L)^2}\bar{q}^{\frac{1}{2}(P^R)^2}\nonumber
\end{align}

The special form of this term allows for the following rewriting
\begin{align}
E_1^{\text{dec}}=-2i(2g-2)\int \frac{d^2\tau}{\bar{\eta}^{24}}G_{g+1}\sum_{(P^L,P^R)}\epsilon_{ac}v_1^c\frac{\partial}{\partial\tau}\left[\tau_2^{2g-1}(\vp)^{2g-3} P_A^Rq^{\frac{1}{2}(P^L)^2}\bar{q}^{\frac{1}{2}(P^R)^2}\right]\nonumber
\end{align}
while a partial integration in $\tau$ finally yields
\begin{align}
E_1^{\text{dec}}=(2g-2)\pi\int \frac{d^2\tau}{\bar{\eta}^{24}}\tau_2^{2g-3}G_{g}\sum_{(P^L,P^R)}\epsilon_{ac}v_1^c(\vp)^{2g-3} P_A^Rq^{\frac{1}{2}(P^L)^2}\bar{q}^{\frac{1}{2}(P^R)^2},\label{result3}
\end{align}
where we have once more made use of (\ref{Ggeq}). We now confront this result with the following expression
\begin{align}
&v_1^av_{\dot{1}}^{\dot{a}}D_{a\ad,A}\mathcal{F}_{g-1}^{\text{dec}}\equiv \dppsa \mathcal{F}_{g-1}^{\text{dec}},
\end{align}
which can be evaluated exactly in the same way as (\ref{left6})
\begin{align}
&\dppsa\int \frac{d^2\tau}{\bar{\eta}^{24}}\tau_2^{2g-4} G_{g}(\tau,\bar{\tau})\sum_{(P^L,P^R)}\left(\vp\right)^{2g-4}q^{\frac{1}{2}(P^L)^2}\bar{q}^{\frac{1}{2}(P^R)^2}=\nonumber\\
&=-2\pi\int \frac{d^2\tau}{\bar{\eta}^{24}}\tau_2^{2g-3}G_{g}\sum_{(P^L,P^R)}(\vp)^{2g-3}P^R_Aq^{\frac{1}{2}(P^L)^2}\bar{q}^{\frac{1}{2}(P^R)^2}.\nonumber
\end{align}
Comparing this expression to (\ref{result3}), we conclude
\begin{align}
\epsilon^{\dot{a}\dot{b}}\frac{\partial}{\partial v_{\dot{1}}^{\dot{a}}}D_{a\bd,A}\mathcal{F}_g^{\text{dec}}=-\frac{1}{2}(2g-2)\epsilon_{ab}v_1^b\dppsa \mathcal{F}_{g-1}^{\text{dec}}.\label{decom1}
\end{align}
Finally replacing the differentiation with respect to $v_{\dot{1}}^{\ad}$ by one with respect to $v_1^a$, we can derive a similar equation
\begin{align}
\epsilon^{ab}\frac{\partial}{\partial v_{1}^{a}}D_{b\ad,A}\mathcal{F}_g^{\text{dec}}=-\frac{1}{2}(2g-2)\epsilon_{\ad\bd}v_{\dot{1}}^{\bd}\dppsa \mathcal{F}_{g-1}^{\text{dec}}.\label{decom2}
\end{align}
For both equations (\ref{decom1}) and (\ref{decom2}), the same considerations as in the four-dimensional case imply that the right hand side can be interpreted as an anomaly.

Notice that the left-hand side of (\ref{decom1}) and (\ref{decom2}) is exactly the harmonicity condition first derived in \cite{Berkovits:1994vy}. There, however, corrections to the equation by boundary terms of the Riemann surface as well as by certain contact terms in operator product expansions were neglected. In the later work \cite{Ooguri:1995cp}, the missing of these extra contributions was pointed out and it was suggested that an additional contraction with harmonic coordinates would project out all extra terms. This was demonstrated by a careful analysis in the topological twisted theory. Indeed, if we project the free indices of (\ref{decom1}) and (\ref{decom2}) with $v_1$ and $v_{\dot{1}}$ respectively, we find
\begin{align}
v_1^a\epsilon^{\dot{a}\dot{b}}\frac{\partial}{\partial v_{\dot{1}}^{\dot{a}}}D_{a\bd,A}\mathcal{F}_g^{\text{dec}}=-\frac{1}{2}(2g-2)v_1^a\epsilon_{ab}v_1^b\dppsa\mathcal{F}_{g-1}^{\text{dec}}=0,\nonumber\\
v_{\dot{1}}^{\ad}\epsilon^{ab}\frac{\partial}{\partial v_{1}^{a}}D_{b\ad,A}\mathcal{F}_g^{\text{dec}}=-\frac{1}{2}(2g-2)v_{\dot{1}}^{\ad}\epsilon_{\ad\bd}v_{\dot{1}}^{\bd}\dppsa\mathcal{F}_{g-1}^{\text{dec}}=0,\nonumber
\end{align}
in complete agreement with \cite{Ooguri:1995cp}, serving as an additional check for our computation.
\subsubsection{Second order relation}
Finally, we can also study the decompactification limit of the second order constraint (\ref{secondorderstring}), whose left-hand side becomes the following differential operator:
\begin{align}
\epsilon^{ab}\epsilon^{\ad\bd}D_{a\ad,A}D_{b\bd,B}\mathcal{F}_g^{\text{dec}}.\label{6dsecondexpression}
\end{align}
Using again the differentiation rules (\ref{naiveRules6D}), we can evaluate (\ref{6dsecondexpression}) in a straight-forward way
\begin{align}
&E_2^{\text{dec}}\equiv\epsilon^{ab}\epsilon^{\ad\bd}D_{a\ad,A}D_{b\bd,B}\mathcal{F}_g^{\text{dec}}=\nonumber\\
&=D_{a\ad,A}\int \frac{d^2\tau}{\bar{\eta}^{24}}\tau_2^{2g-2}G_{g+1}\sum_{(P^L,P^R)}(\vp)^{2g-3}\left[(2g-2)v_1^a\vd^{\ad}-2\pi\tau_2\epsilon^{ab}\epsilon^{\ad\bd}(\vp)P^L_{b\bd}\right]\cdot\nonumber\\
&\hspace{0.8cm}\cdot P^R_Bq^{\frac{1}{2}(P^L)^2}\bar{q}^{\frac{1}{2}(P^R)^2}=\nonumber\\
&=-4\pi \int \frac{d^2\tau}{\bar{\eta}^{24}}\tau_2^{2g-2}G_{g+1}\sum_{(P^L,P^R)}\left[2g\tau_2-2\pi\tau_2^2(P^L)^2\right]P^R_AP^R_B(\vp)^{2g-2}q^{\frac{1}{2}(P^L)^2}\bar{q}^{\frac{1}{2}(P^R)^2}+\nonumber\\
&\hspace{0.3cm}+\delta_{AB}\int\frac{d^2\tau}{\bar{\eta}^{24}}\tau_2^{2g-2}G_{g+1}\sum_{(P^L,P^R)}\left[(2g-1)-2\pi\tau_2(P^L)^2\right](\vp)^{2g-2}q^{\frac{1}{2}(P^L)^2}\bar{q}^{\frac{1}{2}(P^R)^2}-\nonumber\\
&\hspace{0.3cm}-g\delta_{AB}\mathcal{F}_g^{\text{dec}}.\nonumber
\end{align}
Following the same steps as before, we can re-write the first two lines as total derivatives with respect to $\tau$, namely
\begin{align}
E_2^{\text{dec}}&=8\pi i\int\frac{d^2\tau}{\bar{\eta}^{24}}G_{g+1}\sum_{(P^L,P^R)}\frac{\partial}{\partial\tau}\left[\tau_2^{2g}(\vp)^{2g-2}P^R_AP^R_Bq^{\frac{1}{2}(P^L)^2}\bar{q}^{\frac{1}{2}(P^R)^2}\right]+\nonumber\\
&\hspace{0.3cm}+2i\delta_{AB}\int\frac{d^2\tau}{\bar{\eta}^{24}}G_{g+1}\sum_{(P^L,P^R)}\frac{\partial}{\partial\tau}\left[\tau_2^{2g-1}(\vp)^{2g-2}q^{\frac{1}{2}(P^L)^2}\bar{q}^{\frac{1}{2}(P^R)^2}\right]-\nonumber\\
&\hspace{0.3cm}-g\delta_{AB}\mathcal{F}_g^{\text{dec}},\nonumber
\end{align}
which after a partial integration become
\begin{align}
E_2^{\text{dec}}&=4\pi^2\int\frac{d^2\tau}{\bar{\eta}^{24}}\tau_2^{2g-2}G_{g}\sum_{(P^L,P^R)}(\vp)^{2g-2}P^R_AP^R_Bq^{\frac{1}{2}(P^L)^2}\bar{q}^{\frac{1}{2}(P^R)^2}-\nonumber\\
&\hspace{0.3cm}-\pi\delta_{AB}\int\frac{d^2\tau}{\bar{\eta}^{24}}\tau_2^{2g-3}G_{g}\sum_{(P^L,P^R)}(\vp)^{2g-2}q^{\frac{1}{2}(P^L)^2}\bar{q}^{\frac{1}{2}(P^R)^2}-g\delta_{AB}\mathcal{F}_g^{\text{dec}}.\nonumber
\end{align}
In order to simplify this result, we also evaluate the expression
\begin{align}
&\dppsa\dppsb\mathcal{F}_{g-1}^{\text{dec}}=\nonumber\\
&=-2\pi \dppsa \int \frac{d^2\tau}{\bar{\eta}^{24}}\tau_2^{2g-3}G_g\sum_{(P^L,P^R)}(\vp)^{2g-3}P^R_Bq^{\frac{1}{2}(P^L)^2}\bar{q}^{\frac{1}{2}(P^R)^2}=\nonumber\\
&=4\pi^2\int \frac{d^2\tau}{\bar{\eta}^{24}}\tau_2^{2g-2}G_g\sum_{(P^L,P^R)}(\vp)^{2g-2}P^L_{a\ad}P^R_AP^R_Bq^{\frac{1}{2}(P^L)^2}\bar{q}^{\frac{1}{2}(P^R)^2}-\nonumber\\
&\hspace{0.3cm}-\pi\delta_{AB}\int \frac{d^2\tau}{\bar{\eta}^{24}}\tau_2^{2g-3}G_g\sum_{(P^L,P^R)}(\vp)^{2g-2}P^L_{a\ad}q^{\frac{1}{2}(P^L)^2}\bar{q}^{\frac{1}{2}(P^R)^2}.
\end{align}
We can thus obtain the relation
\begin{align}
\epsilon^{ab}\epsilon^{\ad\bd}D_{a\ad,A}D_{b\bd,B}\mathcal{F}_g^{\text{dec}}=\dppsa\dppsb\mathcal{F}_{g-1}^{\text{dec}}-g\delta_{AB}\mathcal{F}_g^{\text{dec}}. \label{strire}
\end{align}
As already mentioned in Section \ref{Sect:Origin6dimensions}, the general structure of this equation can be anticipated from field theory, especially, the existence of the term proportional to $\delta_{AB}$ on the right hand side of (\ref{strire}). However, due to the lack of the setup of conformal supergravity, we are not in a position to predict the exact coefficient $-g$, which is also different from the coefficient in the four-dimensional analog of the second order constraint (\ref{secondorderstring}).

Note finally, that the six-dimensional couplings $\mathcal{F}^{\text{dec}}_g$ (\ref{fgdec}) of the 1/2-BPS effective operator (\ref{effact6}), although obtained by taking the decompactification limit of the $\cN=4$ topological amplitudes ${\cal F}_g^{(3)}$, they are not given by the topological theory on $K3$. The reason is that in their exact genus $g+1$ type II expression, the $\det{\rm Im}\tau$ factors from the space-time coordinates do not cancel. Thus, these couplings are semi-topological, in the sense that string oscillator modes do not contribute, and upon compactification on a $T^2$ they become exactly topological.

\section{Conclusions}\label{Sec:Concl}
In conclusion, in this work, we generalized the holomorphicity property of the $\cN=2$ supersymmetric couplings involving vector multiplets to the moduli dependence of the $\cN=4$ couplings of 1/2-BPS operators defined in harmonic superspace. An example of such operators is provided by the two series found in \cite{Antoniadis:2006mr}, involving powers of the (superdescendant of the) $\cN=4$ chiral Weyl superfield $K$ whose coupling-coefficients are functions of the $\cN=4$ vector moduli and are computed by the $\cN=4$ topological string on $K3\times T^2$.

The resulting harmonicity or analyticity property is expressed in terms of two sets of differential constraints: the first requires the vanishing of one scalar and one harmonic derivatives, while the second imposes two scalar (covariant) derivatives to give back the same coupling up to a multiplicative constant proportional to its (super)conformal weight. We verified these equations on the string side using the explicit expressions for the couplings of one of the two series as 1-loop heterotic integrals on $T^6$.

We also extended the above analysis to $\cN=2$ 1/2-BPS terms in six dimensions and we checked the resulting equations for the couplings obtained in the decompactification limit of the four-dimensional $\cN=4$ topological amplitudes considered before. In principle, our analysis can be generalized in a straight-forward way to the couplings of any 1/2-BPS operator of extended supersymmetry in any space-time dimension.

\section*{Acknowledgements}
We have profited form enlightening discussions with N. Berkovits, B. de Wit, S. Ferrara and E. Ivanov.
This work was supported in part by the European Commission under the RTN contract MRTN-CT-2004-503369, in part by the INTAS contract 03-51-6346 and in part by the French Agence Nationale de la Recherche, contract ANR-06-BLAN-0142. The work of S.H. was supported by the Austrian Bundesministerium f\"ur Wissenschaft und Forschung.


\end{document}